\documentstyle[preprint,eqsecnum,prb,aps,tighten]{revtex}
\begin{document}
\preprint{UBCTP-94-003,UCSB-TH-94, PUPT-94, cond-mat/9409100}
\draft
\title{Conformal Field Theory Approach to the 2-Impurity Kondo
Problem: Comparison with Numerical Renormalization Group Results
} \author{Ian Affleck$^a$, Andreas W.W. Ludwig$^b$ and Barbara
A. Jones$^c$ }
\address{$^{a}$Canadian Institute for Advanced
Research and Physics Department} \address{University of British
Columbia, Vancouver, BC, V6T 1Z1, Canada} \address{$^b$Physics Department,
University of California, Santa Barbara, CA93102}
 \address{$^c$IBM Almaden Research Center, 650 Harry Road, San Jose,
CA 95120} \date{\today} \maketitle \begin{abstract} Numerical
renormalization group and conformal field theory work indicate that
the two impurity Kondo Hamiltonian has a non-Fermi liquid critical
point separating the Kondo-screening phase from the inter-impurity
singlet phase when particle-hole (P-H) symmetry is maintained.  We clarify
the circumstances under which this critical point occurs, pointing out
that there are two types of P-H symmetry.  Only one of them guarantees the
occurance of the critical point.  Much of the previous numerical work was
done on models with the other type of P-H symmetry.  We analyse this
critical point using the boundary conformal field theory technique.  The
finite-size spectrum is presented in detail and compared with about 50
energy levels obtained using the numerical renormalization group. Various
Green's functions, general renormalization group behaviour, and a
hidden $SO(7)$ are analysed.   \end{abstract} 
\section{Introduction} While the single-impurity Kondo problem is by
now rather well understood,\cite{Noz,Bethe,AL2ch} the Kondo lattice model,
of possible relevance to heavy fermion materials, presents additional
difficulties.\cite{Lee}  In particular there are two competing
tendencies; the Kondo effect leads to a magnetic screening of the
spins by conduction electrons while the RKKY interactions between the
spins may lead to antiferromagnetism. The two-impurity Kondo
problem\cite{Jay,Jones,Fye} provides a simple model in which to study
these competing effects.  Of course, in this model, true
antiferromagnetism with static moments (i.e. spontaneous breaking of
rotational symmetry) cannot occur.  Rather the RKKY couplings tend to
promote singlet formation by the two impurities.

A very general form of the Hamiltonian has two $s=1/2$
spins, symmetrically located about the origin, interacting with a Fermi
gas.   The Hamiltonian is: \begin{equation}
H-\mu N=H_0 + H_K +H_{\hbox{self}}\end{equation} where $H_0$, $H_K$ and
$H_{\hbox{self}}$ are the kinetic energy, Kondo interaction and impurity
self interactions respectively: \begin{eqnarray} H_0 &=& \int d^3\vec k
\epsilon (\vec k)\psi^{\alpha \dagger}_{\vec k}\psi_{\alpha \vec k}\nonumber
\\ H_K &=&\int d^3\vec k_1\int d^3\vec k_2\psi^{\alpha
\dagger}_{\vec k_1}\vec \sigma^{\beta}_{\alpha}\psi_{\beta \vec
k_2}\cdot\left[v(\vec k_1)^*v(\vec k_2) \vec S_1 +  v(-\vec k_1)^*v(-\vec
k_2)\vec S_2\right]\nonumber \\ H_{\hbox{self}} &=& K\vec S_1\cdot \vec S_2
\label{Ham3D}\end{eqnarray} $v(\vec k )$ is related to an Anderson model
hybridization matrix element and $K$ is the inter-impurity coupling.

The physics of this model becomes simple in the limits $K \to \pm
\infty$.  When $K \to +\infty$ the two impurities lock into a
singlet state.  Therefore the Kondo coupling has no effect and
the electron gas is completely unaffected by the impurities.
Conversely, when $K \to -\infty$, the impurities become an
effective single $s=1$ impurity.  As we shall see in the next
section, the extended nature of this $s=1$ impurity implies that
two ``channels'' of conduction electrons interact with it.  The
two channel, $s=1$ Kondo problem was studied by Nozi\`eres and
Blandin.\cite{NB} They concluded that, at low temperatures, the
impurity gets magnetically screened by the conduction electrons.
The remaining low energy conduction electron degrees of freedom
are decoupled from the impurity but experience a $\pi \over 2$
phase shift in both channels.  This corresponds to a local Fermi
liquid fixed point.  The low energy electronic degrees of freedom
are free electron-like; the many body interactions induced by the
Kondo interaction lead to a simple phase shift at low energies.

The behaviour of the system at intermediate values of $K$ is less
obvious.  In general, one might expect that a local Fermi liquid
description holds at low temperatures for all $K$.  The phase
shifts could vary continuously with $K$.  However, it was argued
 by Millis, Kotliar and Jones\cite{Millis} that this cannot
happen if P-H symmetry is maintained while $K$ is
varied.  In this case, the zero-energy phase shift, if it is
well-defined, can only be $0$ or $\pi /2$.  Since the $K\to \pm
\infty$ fixed points are stable, it follows that there must be at least
one point in the phase diagram not belonging to either phase,
corresponding to some sort of phase transition.  Such a transition could, in
principle, be first order. Alternatively, if it is continuous, the critical
point cannot be characterizable by  phase shifts.  Numerical renormalization
group (NRG) work indicates\cite{Jones} that the transition is indeed
continuous. The critical point is of non-Fermi liquid type. On the other
hand, quantum Monte Carlo (QMC) work has not seen such a critical
point.\cite{Fye} NRG work on the Anderson model\cite{Sakai} has seen the
critical point only in the case of ``energy-independent coupling
constants''.  In both cases P-H symmetry was maintained.

In the next section we re-examine the
argument for a critical point in more detail.  We show that there are
actually two quite different types of P-H symmetry which may occur in
models of this type; this is connected with the parity symmetry of the
models.  Only one of these P-H symmetries guarantees a phase
transition.  Much of the numerical work in Refs.
(\onlinecite{Fye,Sakai}) was done using models with the ``wrong'' type
of P-H symmetry.

Two of us have recently developed a new boundary conformal field
theory technique to study such non-trivial critical
points.\cite{AL,AL2ch,ALg} The essence of this method is that the
local interactions can be replaced by a conformally invariant
boundary conditions\cite{Cardy} in the low energy effective
Hamiltonian.  One of the main purposes of this paper is to explain the
application of the technique to this problem and elucidate its
various predictions.  A brief presentation of some of these
results was made earlier.\cite{AL2imp}  In Sec. III we review the
reduction of the problem to a one-dimensional field theory and then
discuss a convenient and somewhat unconventional ``bosonization'' of the
model in terms of Wess-Zumino-Witten matrix fields and Ising model
fields.  In Section IV we solve for the non-trivial critical point by
hypothesizing
 the corresponding boundary conditions (there are only
a few  possibilities).  We calculate the
resulting finite size spectrum using this boundary condition and
compare it to the NRG results, obtaining excellent agreement.  In
Sec.  V we discuss the various scaling operators at the
non-trivial critical point, the stability of the fixed point (including
the effects of P-H symmetry breaking) and the behaviour of various Green's
functions.  The non-trivial fixed point possesses a remarkable hidden
$SO(7)$ symmetry.\cite{AL2imp}  We explain this in detail in Sec. VI and
comment on the possible connection with a physical picture
of the 2-impurity Kondo problem based on abelian bosonization
and ``refermionization''.\cite{Sire,Gan}  In Sec. VII we summarize the
disagreement with other calculations and the prospects for resolving them.

\section{Existence of the Non-trivial Fixed Point and
Particle-Hole Symmetry}

We begin with a few comments on the Hamiltonian of Eq. (\ref{Ham3D}).
  Often, one does not consider an explicit inter-impurity interaction but
only the indirect RKKY interaction, of second and higher order in $J$.  More
generally an additional interaction could be generated by other exchange
processes not involving the conduction electrons.  Often one considers a
$\delta$-function Kondo interaction, with the impurities at $\pm \vec
R /2$ in which case:
\begin{equation} v(\vec k )=v_0e^{i\vec k \cdot \vec
R/2}.\label{delta}\end{equation}
 The more
general form that we consider allows for longer range hybridization.
Note that, if $\epsilon (-\vec k)=\epsilon (\vec k)$, the Hamiltonian is
invariant under the parity transformation: \begin{eqnarray} \vec
S_1&\leftrightarrow &\vec S_2\nonumber \\ \psi_{\vec k} &\leftrightarrow &
\psi_{-\vec k},\label{par_def}\end{eqnarray} corresponding to a reflection
about a point in position space (the midpoint between the 2 impurities.)
For a tight-binding model, the origin of parity symmetry may not correspond
to a lattice point.  Nonetheless, we define our Fourier transforms with
respect to it.

For certain choices of the dispersion relation, $\epsilon (\vec k)$
and hybridization matrix element, $v(\vec k )$, the Hamiltonian of Eq.
(\ref{Ham3D}) will have particle-hole (P-H) symmetry.  We consider invariance
of $H$ under a general P-H transformation of the form:
\begin{eqnarray}\psi_{\alpha \vec k}&\to& \epsilon_{\alpha
\beta}\psi^{\beta \dagger}_{\vec k'}\nonumber
\\
\vec S_1&\to& \vec S_1\nonumber \\
\vec S_2&\to& \vec S_2.\label{3D_PH0}\end{eqnarray} where $\vec k'$ is some
function of $\vec k$.  Invariance of $H_0$ requires: \begin{equation}
\epsilon (\vec k)=-\epsilon (\vec k').\label{PHep}\end{equation}
Invariance of the Kondo interaction requires (after a possible phase
redefinition of $v$ which doesn't effect $H_K$): \begin{eqnarray} v(\vec
k')&=&v(\vec k)^*\nonumber \\ v(-\vec k' )&=&v(-\vec k)^*e^{i\alpha}
\label{3D_PH}\end{eqnarray} where $\alpha$ is a $\vec k$-independent
phase.  For instance, for a cubic lattice tight-binding model at
half-filling, \begin{equation} \vec k' = \vec k_0-\vec k,\end{equation}
where $\vec k_0$ is the nesting wave-vector: \begin{equation} \vec
k_0\equiv (\pi /a,\pi /a,\pi /a).\end{equation}Hence, for the
$\delta$-function Kondo interaction of Eq. (\ref{delta}), and appropriate
choice of the phase of $v_0$; Eq. (\ref{3D_PH}) is obeyed with:
\begin{equation} \alpha = \vec k_0\cdot \vec R.\label{alpha}\end{equation}
Assuming that $\vec R$ (the vector connecting the two impurities) is  a
lattice vector, the phase $\alpha$ can take on the values $0$ or $\pi$.

When particle-hole symmetry is present, there is also an exact
$SU(2)$ ``isospin'' symmetry, in addition to the normal spin
symmetry.\cite{Jones}  [For a general discussion of this see Sec, II of
Ref. (\onlinecite{ALPC}).]  One of these isospin symmetry generators is
simply the total electron number, corresponding to the $z$-component of
isospin, $I^z$.  The lowering operator, $I^-$ is proportional to:
\begin{equation} \int d^3\vec k \psi_{\vec k,\uparrow}\psi_{\vec
k',\downarrow},\end{equation} where $\vec k'$ is the function of
$\vec k$  occuring in the particle-hole transformation.

The Hamiltonian of Eq. (\ref{Ham3D}) can be reduced {\it exactly} to a
one-dimensional one.  To do this we define two fields:
\begin{equation} \psi_{\pm ,E}\equiv \int d^3\vec k \delta [\epsilon
(\vec k )-E]v(\pm \vec k)\psi_{\vec
k}.\label{psi+-def}\end{equation}  Note that, of the infinite set of fields
of a given energy $E$, (corresponding to all points on the constant energy
surface in $\vec k$-space) only these two appear in the Kondo
interaction.  We may define parity even and odd orthonormal linear
combinations of these two fields:
\begin{eqnarray}\psi_{e,E}&\equiv &
(\psi_{+,E}+\psi_{-,E})/N_e(E)\nonumber \\ \psi_{o,E}&\equiv
&(\psi_{+,E}-\psi_{-,E})/N_o(E),\label{eodef}\end{eqnarray} where:
\begin{equation} N_{e,o}(E) \equiv \int d^3\vec k\delta [E-\epsilon
(\vec k)]|v(\vec k )\pm v(-\vec k )|^2.\label{Neodef}\end{equation}
The anti-commutators are normalized as follows:
\begin{equation} \{\psi_E,\psi^\dagger_{E'}\}=\delta (E-E').\end{equation}
We may complete $\psi_e$ and $\psi_o$ into a complete orthonormal
basis for each energy, $E$; only these two fields appear in the Kondo
interaction.  Discarding the additional fields which decouple, we
obtain an exact one-dimensional  rewriting of the original
Hamiltonian, with:
\begin{eqnarray}H_0& =&\int dE
E[\psi^\dagger_{e,E}\psi_{e,E}+\psi^\dagger_{o,E}\psi_{o,E}]\nonumber
\\ H_K&=&
\int dE dE'\bigg\{\left( N_e(E)N_e(E')\psi^{\dagger}_{e,E}\vec
\sigma \psi_{e,E'} +N_o(E)N_o(E')\psi^{\dagger}_{o,E}\vec \sigma
\psi_{o,E'}\right) \cdot (\vec S_1+\vec S_2) \nonumber \\
&&+N_e(E)N_o(E')\left(\psi^{\dagger}_{e,E}\vec \sigma
\psi_{o,E'}+\psi^{\dagger}_{o,E}\vec \sigma \psi_{e,E'}\right)\cdot
(\vec S_1-\vec S_2) \biggr\}.\label{H1D}\end{eqnarray}

If the original problem had the particle-hole symmetry of Eq.
(\ref{3D_PH}), then the one-dimensional problem also has P-H symmetry.
The transformation of the fields $\psi_{e,E}$ and $\psi_{o,E}$ can be
deduced from:
\begin{eqnarray} \psi_{+ ,E} &\to & \psi_{+
,-E}^\dagger \nonumber \\
\psi_{- ,E} &\to & e^{i\alpha}\psi_{-
,-E}^\dagger ,\label{pmPH}\end{eqnarray} which in turn follows from Eq.
(\ref{psi+-def}) using Eq. (\ref{3D_PH0})-(\ref{3D_PH}).

[We suppress the transformation of spin indices; it is the same as in Eq.
(\ref{3D_PH0}).] In the case $\alpha = 0$ we have [ using Eq. (\ref{PHep})
and (\ref{3D_PH})]:
\begin{eqnarray}N_{e}(-E)&=&N_{e}(E)\nonumber \\
N_{o}(-E)&=&N_{o}(E)\label{N0}\end{eqnarray} and
\begin{eqnarray} \psi_{e,E}&\to & \psi_{e,-E}^\dagger \nonumber \\
\psi_{o,E}&\to &\psi_{o,-E}^\dagger \label{PH1D0} .\end{eqnarray} For
$\alpha =\pi $, we find: \begin{equation} N_e(-E) =
N_o(E),\label{Npi}\end{equation} and \begin{eqnarray} \psi_{e,E}\to
\psi_{o,-E}^\dagger \nonumber \\ \psi_{o,E}\to \psi_{e,-E}^\dagger
.\label{PH1Dpi}\end{eqnarray}   It can be directly verified that the P-H
transformation of Eq. (\ref{PH1D0}) or (\ref{PH1Dpi}) are symmetries of the
one-dimensional Hamiltonian of Eq. (\ref{H1D}) if $N_e(E)$ and $N_o(E)$
satisfy Eq. (\ref{N0}) or (\ref{Npi}) respectively.

It turns out that the model behaves very differently depending on
which type of P-H symmetry is present. With the first type, a phase
transition must separate the Kondo-screened and inter-impurity
singlet phases; with the second type no transition is required
or expected to occur. We reiterate the argument for a transition with
the first type of P-H symmetry.\cite{Millis}  In a Fermi liquid phase,
it should be possible to characterize the zero-temperature fixed point
by phase shifts for the even and odd channels at $E=0$.  These amount to
boundary conditions relating incoming and outgoing operators:
\begin{eqnarray}\psi_{e,E}^{\hbox{out}}&=&e^{2i\delta_{e}}
\psi_{e,E}^{\hbox{in}}\nonumber \\
\psi_{o,E}^{\hbox{out}}&=&e^{2i\delta_{o}}
\psi_{o,E}^{\hbox{in}}.\label{bc}\end{eqnarray}
  The Hermitean conjugate fields, $\psi^\dagger_{e,o}$
obey the same conditions with $\delta_{e,o}\to -\delta_{e,o}$.  Hence,
if the first type of P-H symmetry is obeyed, both $\delta_e$ and
$\delta_o$ can only take the values $0$ or $\pi/2$ [from Eq.
(\ref{PH1D0}) at $E=0$, right at the Fermi energy]. [Note that the boundary
conditions of Eq. (\ref{bc}) only depend on $2\delta$ (mod $2\pi$) so that
$\delta =0$ or $\pi$ are equivalent as are $\delta =\pm \pi /2$.]
$\delta_e=\delta_o=0$ in the inter-impurity singlet phase and
$\delta_e=\delta_o=\pi/2$ in the Kondo-screening phase. Furthermore, both
of these fixed points are absolutely stable (no relevant {\it or marginal}
operators.)  It then follows that there must be some sort of phase
transition separating these two phases. It could be first order, or
correspond to a non-Fermi liquid critical point at which the phase shifts
are not defined.  On the other hand, if we have the second type of P-H
symmetry, $\delta_e$ and $\delta_o$ may take on arbitrary values subject
only to the condition $\delta_e=-\delta_o$.  Thus they may vary
continuously from $(0,0)$ in the limit of $\infty$ antiferromagnetic
inter-impurity coupling to $(\pi /2,-\pi /2)$ in the opposite limit of
$\infty$ ferromagnetic inter-impurity coupling.  No transition is
necessary in this case.

The same argument can be made by considering possible potential
scattering terms in the Hamiltonian.  While the original Hamiltonian
has no such terms, they will be generated in higher orders of
perturbation theory.  In general these take the form: \begin{equation}
H_{PS} = \int dE
dE'[V_e(E,E')\psi^\dagger_{e,E}\psi_{e,E'}+V_o(E,E')\psi^\dagger_{o,E}
\psi_{o,E'}].\end{equation}  The phase shifts at zero energy contain
terms proportional to $V_e(0,0)$ and $V_o(0,0)$ respectively, for weak
potential scattering, ignoring the Kondo interactions.  With the first
type of P-H symmetry:
\begin{equation} V_{e,o}(E,E')=-V_{e,o}(-E',-E).\end{equation}  This
implies $V_{e,o}(0,0)=0$; no phase shift at zero energy.  On the other
hand the second type of P-H symmetry implies:
\begin{equation} V_{e}(E,E')=-V_{o}(-E',-E).\end{equation} At zero
energy this gives: $V_e(0,0)=-V_o(0,0)$, allowing (equal and opposite)
phase shifts at zero energy.

In Sec. IV we explicitly study the stability of the non-trivial
critical point under potential scattering.  Our analysis shows that
 $V_e(0,0)=-V_o(0,0)\neq 0$ is a relevant perturbation.

\section{One Dimensional Bose-Ising Representation}
In this section we derive a representation of the two impurity
Kondo problem which is convenient for studying the non-trivial
critical point. The one-dimensional Hamiltonian of Eq. (\ref{H1D}) is
treated using a generalized bosonization method based on the
Goddard-Kent-Olive\cite{GKO} coset construction.  We end up
representing the fermions by three Wess-Zumino-Witten non-linear
$\sigma$-models\cite{WZW} together with an Ising model.

Taylor-expanding  $N_{e,o}(E)$ in Eq. (\ref{H1D}) around $E=0$, we obtain
the non-derivative interactions:
\begin{eqnarray} H_K&=&{1\over 2}
\int dE dE'\bigg\{ \left(J_e\psi^{\dagger}_{e,E}\vec \sigma
\psi_{e,E'} +J_o\psi^{\dagger}_{o,E}\vec \sigma
\psi_{o,E'}\right) \cdot (\vec S_1+\vec S_2) \nonumber \\
&&+J_m\left(\psi^{\dagger}_{e,E}\vec \sigma
\psi_{o,E'}+\psi^{\dagger}_{o,E}\vec \sigma \psi_{e,E'}\right)\cdot
(\vec S_1-\vec S_2) \biggr\}\end{eqnarray} where \begin{equation}J_e\equiv
2N_e(0)^2, \ \  J_o\equiv 2N_o(0)^2, \ \ J_m\equiv 2N_o(0)N_o(0).
\end{equation}  Sums over spin indices are implied.  We effectively obtain
a two channel one-dimensional model.   The origin of the two channels
is simply the fact that the two impurities are at different
spatial locations.
In general, with $n$ impurities at different
locations, we would obtain an $n$-channel one-dimensional model
at long wavelengths.

 The higher terms in the Taylor expansion
of $N_{e,o}(E)$ give various derivative interactions.  These are irrelevant
at the weak coupling (unstable) critical point, assuming the first type of
particle-hole symmetry discussed in Sec. II.  That is to say, they are
irrelevant assuming they don't generate any potential scattering terms; this
is the case with the first type of P-H symmetry.  From our analysis of the
non-trivial critical point we will conclude that these extra terms are also
irrelevant at that critical point, assuming the first type of P-H symmetry.

We now see that, in the limit of an infinite {\it ferromagnetic}
inter-impurity interaction $K$, the model reduces to a
single-impurity two-channel $s=1$ Kondo problem, with different
couplings, $J_e$ and $J_o$ to the two channels.  This difference
is known to be irrelevant\cite{NB}.

For analysing the critical point, we actually find it somewhat
more convenient to introduce a different orthonormal basis:
\begin{equation} \psi_{1,2} \equiv (\psi_e \pm
\psi_o)/\sqrt{2}\label{flavdef}\end{equation} The kinetic energy
remains diagonal, as in Eq. (\ref{H1D}) and the Kondo interaction
becomes: \begin{eqnarray} H_K &=& {1\over 2}\int dE dE'
\{J_+[\psi^{\dagger}_{1,E}\vec \sigma
\psi_{1,E'}+\psi^{\dagger}_{2,E}\vec \sigma \psi_{2,E'}]\cdot
[\vec S_1+\vec S_2] +J_m[\psi^{\dagger}_{1,E}\vec \sigma
\psi_{1,E'} -\nonumber \\ &&\psi^{\dagger}_{2,E}\vec \sigma
\psi_{2,E'}]\cdot [\vec S_1-\vec S_2]
+J_-[\psi^{\dagger}_{1,E}\vec \sigma
\psi_{2,E'}+\psi^{\dagger}_{2,E}\vec \sigma \psi_{1,E'}]\cdot [\vec
S_1+\vec S_2]\}\label{H_K}\end{eqnarray} where \begin{equation}
J_{\pm} \equiv (J_e\pm J_o)/2\label{H_K12}\end{equation} Note that
$\psi_{1,2}$ are not, in general, the same as $\psi_{\pm}$ since
$N_e\neq N_o$.  We emphasize that this is the most general
Hamiltonian consistent with particle-hole symmetry, up to
operators which are irrelevant  at the (unstable) zero-coupling
fixed point, ie. in ordinary perturbation theory.

The conserved isospin operators referred to above take simple
forms in the effective one-dimensional theory:
\begin{eqnarray} I^z &\equiv&
{1\over 2}\int dE [\psi^{\dagger \alpha}_{ 1,E}\psi_{ \alpha
1,E}+{1\over 2}\psi^{\dagger \alpha}_{ 2,E}\psi_{ \alpha
2,E}]\nonumber \\ I^- &\equiv& \int dE [\psi_{ \uparrow 1,E}\psi_{
\downarrow 1,-E}+\psi_{ \uparrow 2,E}\psi_{\downarrow 2,-E}]
\label{isospin} \end{eqnarray}  We note that when $J_-=0$ there is even more
symmetry.  Now the charges of the $1$ and $2$ species of fermions
are separately conserved and in fact we have two commuting
sets of isospin generators:
\begin{eqnarray} I^z_{ 1} &\equiv&
{1\over 2}\int dE \psi^{\dagger \alpha}_{ 1,E}\psi_{ \alpha
1,E}\nonumber \\ I^-_{ 1} &\equiv& \int dE\psi_{ \uparrow 1,E}\psi_{
\downarrow 1,-E},\label{isospin_1} \end{eqnarray}and similarly for
$\vec I_2$.  While the isospin generators $\vec I_1$ and $\vec I_2$, obey
the usual $SU(2)$ commutation relations, they commute with each other and
also with the normal spin rotation generators, $\vec J$ (not to be confused
with the coupling constants).

In position space the
one-dimensional problem can be defined in terms of left-movers
only: \begin{equation} \psi_L(t,x) = {1\over \sqrt{v}}\int_{-D}^{D} dE
e^{-iE(t+x/v)}\psi_{E},\label{psiL} \end{equation} for
$-\infty <x <\infty$.  Here $D$ is a bandwidth cut-off and $v$ is an
arbitrary velocity parameter which defines the scale of length in the
effective one-dimensional problem. We adopt an unconventional normalization
for our one-dimensional  position space fermion fields so that they obey:
\begin{equation} \{\psi_L^{\dagger}(x),\psi_L(y)\}=2\pi \delta
(x-y)\end{equation} Alternatively, we may define right and left-movers on
the half-line, $x>0$, with  \begin{equation} \psi_R(t,x) \equiv
\psi_L(t,-x), \ \ (x>0)\label{psiR}\end{equation} and $\psi_L(x,t)$ defined
as in Eq. (\ref{psiL}).  Note that Eq. (\ref{psiR}) would follow
automatically from defining $\psi_R(x,t)$ to be the right-moving field
obeying the boundary condition $\psi_R(t,0)=\psi_L(t,0)$.  (Henceforth we
drop $L$, $R$, subscripts.  Fields are generally left-movers.)

A key feature in the conformal field theory approach to this
problem is bosonization. Actually, we use a somewhat generalized
version of bosonization in which fermion fields are represented
in terms of boson fields and Ising fields. A conventional abelian
bosonization approach would involve introducing four bosons for
the spin and charge degrees of freedom of each channel,
$\phi_{s,i}$, $\phi_{c,i}$ with $i=1,2$.  In fact we only
introduce the left-moving components of these bosons, or
equivalently introduce left and right movers on the half-line
with an appropriate boundary condition. The bosonization formulas
take the form: \begin{equation} \psi_{\uparrow i}\propto
e^{i\sqrt{2\pi}(\phi_{c,i}+\phi_{s,i})}, \ \  \psi_{\downarrow
i}\propto e^{i\sqrt{2\pi}(\phi_{c,i}-\phi_{s,i})}
\label{bos}\end{equation} As discussed elsewhere,\cite{LesH}
these bosons should be regarded as periodic variables since only
their exponentials (or derivatives) occur in physical
quantities.  Adopting a fixed normalization for the free boson
Lagrangian density: \begin{equation} {\cal L} = {1\over
2}\partial_{\mu}\phi\partial^{\mu}\phi\end{equation} we define a
``compactification radius'' by identifying: \begin{equation} \phi
\equiv \phi + 2\pi R\end{equation} The above bosonization
formulas, Eq. (\ref{bos}), identify the compactification radius
for the spin boson as $R=1/\sqrt{2\pi}$.  At this radius, the
free spin boson theory exhibits $SU(2)$ symmetry and is, in fact
equivalent to an $SU(2)$ Wess-Zumino-Witten (WZW) non-linear
$\sigma$ model with Kac-Moody central charge $k=1$.  The spin
factors occurring in Eq. (\ref{bos}) can be written in terms of
the left-moving factors of the WZW fields, $g_{\alpha i}$:
\begin{equation} \psi_{\alpha i} \propto
e^{i\sqrt{2\pi}\phi_{c,i}}g_{\alpha i}\end{equation}

In fact, the charge bosons have the same compactification
radius.  This is no accident.  It reflects the fact that the free
fermion theories for each channel have an $SU(2) \times SU(2)$
 symmetry. This corresponds to separate spin and isospin symmetry
[see Eq. (\ref{isospin_1})] for each channel. This is equivalent to
$O(4)$.  The $O(4)$ symmetry is manifest if the two spin components
are each written in terms of hermitean and anti-hermitean parts,
giving a total of four herimitean fermion fields, for each channel.
The Kondo interactions will break the separate spin symmetries down
to the diagonal subgroup; they have a similar effect on the separate
isospin symmetries except in the special case, $J_-=0$. Thus we may
introduce two more $k=1$ WZW field, $(h_i)_A$ corresponding to the
two charge fields.  Here $A$ is an isospin index whereas $i$ labels
the two species. The bosonization formulas of Eq. (\ref{bos}) then
take the form: \begin{equation} \psi_{\alpha i} \propto
(h_i)_1g_{\alpha i}\end{equation} The left-moving $k=1$ WZW fields
obey: \begin{equation} g^{\alpha \dagger} = \epsilon^{\alpha
\beta}g_{\beta}\label{daggerident}\end{equation} and similarly for
the $h_i$'s, as can be seen by comparing the abelian and non-abelian
bosonization formulas.  As we shall see, the charge bosons play a
purely passive spectator role in the 2-impurity Kondo effect and it
is not necessary to replace the charge bosons by the non-abelian
$h$-fields in what follows.

The next crucial step is to rewrite the bosonized theory in terms
of a total spin boson and some leftover degrees of freedom,
describing relative spin fluctuations.  The obvious way of doing
this, {\it which is not the one we use}, consists of changing
variables to the sum and difference of the spin bosons,
$\phi_{s,1}$ and $\phi_{s,2}$.  The reason that we do not follow
this procedure is that it does not explicitly maintain the total
$SU(2)$ symmetry of ordinary spin.   Instead we use a procedure
based on the Sugawara form of the Hamiltonian.

Before explaining this in detail, we pause to review some basic
properties of conformal field theories and Sugawara
Hamiltonians.  We work with left-movers only (so that all
operators are functions of $x+vt$ only) and scale a factor of
$v/2\pi$ out of the Hamiltonian for convenience, writing:
\begin{equation} H = {v\over 2\pi}\int_{-l}^ldx{\cal H}
\end{equation} Assuming that ${\cal H}(-l)={\cal H}(l)$, we
define the Fourier transform: \begin{equation} L_n\equiv {l\over
2\pi^2}\int dx e^{in\pi /l}{\cal H}(x) \end{equation} For a
general conformally invariant Hamiltonian, these Fourier modes
generate the Virasoro algebra: \begin{equation} [L_n,L_m] =
(n-m)L_{n+m} + {c\over 12\pi}n(n^2-1)\delta_{n+m,0}\end{equation}
where $c$ is the conformal anomaly parameter.  One class of
conformally invariant theories, relevant to the present
discussion, has a Hamiltonian density quadratic in the currents,
$J^a(x)$ ($a=1,2,3,...n$) of some group of dimension $n$.  In the
present case, only the group $SU(2)$, with $n=3$, is relevant.
The Fourier modes of the currents obey the Kac-Moody algebra:
\begin{equation} [J^a_n,J^b_m] = i\epsilon^{abc}J^c_{n+m}
+{1\over 2}kn\delta_{n+m,0}\end{equation} Here $k$ is the
Kac-Moody central charge (or level).  It must be a positive
integer.  The Hamiltonian then takes the Sugawara form:
\begin{equation} {\cal H}(x) = {1\over 2+k}:\vec J(x) \cdot \vec
J(x):\label{Sug}\end{equation} or, in momentum space:
\begin{equation} L_n = {1\over 2+k}\sum_{-\infty}^{\infty}:\vec
J_{-n}\cdot \vec J_{n+m}:\end{equation}
  It follows from the Kac-Moody algebra that the Sugawara
Hamiltonian obeys the Virasoro algebra with conformal anomaly,
\begin{equation} c = {3\over 2+k}\end{equation} The Hamiltonian
generates time-translation of the currents implying the
additional commutation relations: \begin{equation} [L_n, J_m^a] =
-mJ^a_{n+m}\label{LJ}\end{equation} This fixes the normalization
of the Hamiltonian in Eq. (\ref{Sug}).

We now return to the problem at hand, considering at first the
zero coupling, free theory.  The free theory of two channels of
left-moving fermions with spin can be written in Sugawara form.
Altogether there are four commuting terms in the Hamiltonian,
quadratic in the charge and spin currents for channels $1$ and
$2$.  The charge parts can actually be written more symmetrically
as quadratic forms in the Isospin currents defined in Eq.
(\ref{isospin}) but this is not of great importance for what
follows.  The two sets of commuting spin currents for each
channel obey the Kac-Moody algebra at level $k=1$, and the
associated Hamiltonians, ${\cal H}_{si}$, each have conformal
anomaly $c=1$.  It is natural to rewrite the theory in terms of
the total spin currents:\begin{equation} \vec J (x)\equiv \vec
J_1 (x)+\vec J_2 (x) \end{equation}  These obey the Kac-Moody
algebra with $k=2$.  The associated Hamiltonian, constructed from
$\vec J$ using Eq. (\ref{Sug}) with $k=2$, ${\cal H}_s$, has
$c=3/2$.  A crucial point is that ${\cal H}_{s1}+{\cal H}_{s2}$
and ${\cal H}_s$ both generate time-translation for the total
spin currents.  I.e., they both obey the commutation relations of
Eq. (\ref{LJ}) with the total spin currents.  Since ${\cal H}_s$
is itself quadratic in these total spin currents it follows that
\begin{equation} [{\cal H}_{s1}(x)+{\cal H}_{s2}(x)-{\cal
H}_s(x), {\cal H}_s(y)] = 0\end{equation} Thus the spin
Hamiltonian, ${\cal H}_{s1}+{\cal H}_{s2}$, can be written as a
sum of two commuting pieces, ${\cal H}_s$ and a remainder.  This
is an example of the Goddard-Kent-Olive coset
construction.\cite{GKO}  The remainder is associated with the
coset $[SU(2)\times SU(2)]/SU(2)_D$ in this case. (``D''
represents the diagonal subgroup.)  The value of the conformal
anomaly for the coset Hamiltonian is $c=1+1-3/2 = 1/2$.  The
coset Hamiltonian obeys the Virasoro algebra with this value of
$c$. There is a unique unitary conformal field theory with this
value of $c$, namely the Ising model.

Thus rather than replacing the two spin bosons for each channel
by a total spin boson and a difference boson we replace them by a
$k=2$ Kac-Moody conformal field theory and an Ising model.  The
former theory can be considered to be the left-moving part of a
$k=2$ WZW model.  ie., we may construct the various operators out
of the left-moving factor of a unitary matrix, $g_{\alpha}$.  The
value of $k$ for this matrix field is the coefficient of the
Wess-Zumino topological term in the Lagrangian.  We may now write
down representations for the various operators in the free
fermion theory as products of charge (or isospin) bosons, the
total spin boson, $g_{\alpha}$ and the Ising field.  The $k=2$
WZW model has primary fields\cite{Itz} of spin $j=0$ (identity
operator, $\bf 1$), $j=1/2$ (fundamental field, $g_{\alpha}$) and
$j=1$ (denoted $\vec \phi$.). The $k=1$ WZW model only has the
identity operator and the $j=1/2$ primary, $h_A$. Their scaling
dimension is given by the general formula: \begin{equation} x =
{j(j+1)\over 2+k} \end{equation} There are three primary fields
in the Ising model: the identity operator, ${\bf 1}$,  the Ising
order parameter, $\sigma$ of dimension $x=1/16$ and the energy
operator $\epsilon$ with dimension $x=1/2$.  [These are
dimensions of left-moving factors only.  Thus the order parameter
has total dimension $1/8$ corresponding to a correlation
exponent, $\eta=2x=1/4$.  The energy operator has total dimension
$x=1$ corresponding to a thermal exponent, $\nu=2-x=1$.]  The
fermion field is written in this representation as:
\begin{equation} \psi_{\alpha i} \propto (h_i)_1g_{\alpha}\sigma
\label{Isingbos}\end{equation} Note that the three factors have
dimensions which add up correctly to that of the fermion operator:
$x=1/4+3/16+1/16=1/2$.  This determines uniquely the
representation.  Other operator representations can be determined
using the operator product expansion (OPE).  For the $SU(2)$ WZW
fields, $h_i$ and $g$, the OPE of two primary fields of spin $j$
and $j'$ gives each primary field with spin from $|j-j'|$ up to
the minimum of $j+j'$ and $k-j-j'$.  For the Ising model the OPE
gives: \begin{equation} \sigma \times \sigma \to {\bf 1} +
\epsilon ,\qquad  \sigma \times \epsilon \to \sigma , \qquad
\epsilon \times \epsilon \to {\bf 1} \label{fusion}\end{equation}
This OPE is equivalent to that of the $k=2$ WZW model with the
identification of Ising and WZW primary fields: \begin{equation}
\sigma \leftrightarrow g, \qquad \epsilon \leftrightarrow \vec
\phi\label{equiv}\end{equation} Using the OPE, symmetry
considerations and consistency of scaling dimensions, we can
determine the representation of any operator in the free fermion
theory. An important point is that not all products of isospin,
spin and Ising operators occur in the representation of free
fermions.  When taking the OPE of a product of operators
representing a free fermion operator, only a subset of all
operators in the products of the OPE's for each factor occurs.
This type of representation of free fermions is known as a
conformal embedding. The complete set of products of primary
fields that occurs in the representation of free fermions is
given in Table I.  Note that we only list primary fields.  Other
operators such as $\vec J$ are descendents.

A subtlety arises concerning parity.  Parity takes $\psi_o \to
-\psi_o$ in the one-dimensional theory.
In the alternative basis of Eq. (\ref{flavdef}) it interchanges
$\psi_1$ and $\psi_2$.  It thus follows that it interchanges the
two isospin fields, $h_1$ and $h_2$.  However, that is not the
whole story as we see by considering the operator
$\psi_1^{\dagger}\vec \sigma \psi_1-\psi_2^{\dagger}\vec\sigma
\psi_2$.  This is odd under parity but has the representation
$\vec \phi \epsilon$, independent of the isospin fields.  To
obtain a consistent definition of parity we define it to also
take $\vec \phi \to -\vec \phi$.

In the Bose-Ising representation the Kondo interaction is written
as: \begin{equation} {\cal H}_K=\tilde J_+\vec J(0)\cdot (\vec
S_1+\vec S_2) +\tilde J_m\vec \phi (0) \epsilon (0)\cdot (\vec
S_1-\vec S_2) +\tilde J_-(h_1^\dagger )^A(0)(h_2)_A(0)\vec \phi
(0)\cdot (\vec S_1+\vec S_2)\label{Ham-bos}\end{equation} Here
the three coupling constants, $\tilde J_{\pm}$, $\tilde J_m$ are
proportional to the ones defined in Eq. (\ref{H_K}).  Note that
while $\tilde J_+$ only couples the total spin field, $\tilde
J_m$ also couples the Ising field and $\tilde J_-$ also couples
the charge fields. Since $\tilde J_-$ will turn out to be
irrelevant for the particle-hole symmetric case being considered
here, it follows that the relevant part of the Kondo interaction
only involves the total spin and Ising fields.

We may write the finite-size spectrum in terms of the
``bosonized'' representation.  Here we take the one-dimensional
effective theory and impose convenient boundary conditions on it
on a line of finite length.  Because of the rather unsymmetrical
way that the dimensional reduction takes place [see Sec. II], these boundary
conditions would not arise from any simple or natural ones on the original
three-dimensional problem.  Nevertheless we are interested in considering the
finite system for two reasons.  The first is that the Wilson
numerical renormalization group method essentially studies a
finite one-dimensional system.  We will make detailed comparisons
of the finite-size spectrum with the results of this method.  The
second reason is that there is an intimate connection between the
finite-size spectrum and the operator content (for the infinite
system) which we exploit.  Thus we restrict the left-moving
fermions to the interval: \begin{equation} -l < x < l
\end{equation} and impose the boundary condition: \begin
{equation} \psi_L(l)=-\psi_L(-l)\label{freebc}\end{equation} We
note that in the  formalism with left and right movers on the
positive $x$-axis [see Eq. (\ref{psiR})] the boundary condition
becomes: \begin{equation} \psi_L(l)+\psi_R(l)=0\end{equation} The
free fermion spectrum is obtained by populating the single
fermion momentum eigenstates with momentum $k=(n+1/2)\pi /l$.
Note that with these boundary conditions, the groundstate is
unique with all negative $k$ states filled and positive $k$ states
empty.  ($k$ is measured from $k_F$.)  In general states have
energies (measured from that of the groundstate) of the form
$E=(\pi /l)x$ where $x$ is an integer or half-integer.  There is
a one-to-one correspondence between states and operators with
these boundary conditions; the value of $x$ corresponding to the
dimension of the operator. For instance, $x=1/2$ corresponds to a
single particle or hole state.  It is created from the
groundstate by a single application of the fermion field. Just as
the operators can be written as products of charge, spin and
Ising operators, the states can be written as direct products of
states from each sector.  In each sector, the states are grouped
into several conformal towers with energies of the form $E=(\pi
/l)(x+n)$ where $x$ is fixed and fractional and $n$ is an
integer. The conformal towers are in one-to-one correspondence
with the set of primary fields, with $x$ being the scaling
dimension. The set of products of conformal towers from each
sector which can occur in the free fermion spectrum corresponds
to the set of products of primary fields that occur in
representing free fermion operators.  Thus it can be read off
from Table I.
\section{Finite-size Spectrum} In this section we postulate a
critical theory of the unstable multicritical point which occurs
in the particle-hole symmetric case at a critical value of the
inter-impurity coupling, $K$.  We derive the corresponding
finite-size spectrum and compare it to the results of the NRG.

A fundamental assumption behind our approach is that the critical
point can be described as a conformally invariant boundary
condition.\cite{AL}  We expect this to be generally true for a
wide class of critical phenomena involving quantum impurities.
Indeed this assumption is very analogous to the one which is
widely made in studying {\it bulk} critical phenomena in
two-dimensional [or (1+1)-dimensional] systems.  There it is
assumed that a wide class of critical phenomena exhibits
conformal invariance.  Although such an assumption can rarely be
proven, it is consistent in the following sense.

We first of all assume scale invariance at the critical point.
This could be taken as a definition of criticality.  We then
argue that the critical theory should have $SO(2)$ or Lorentz
[$SL(1,1)$] invariance.  Although this is not a symmetry of most
underlying microscopic theories it can be seen that the operators
which break this symmetry down to whatever subgroup exists in the
microscopic theory are irrelevant.  In the case of a
two-dimensional theory defined on the square lattice, the
subgroup of $SO(2)$ is the symmetry group of the square lattice.
This permits non-rotationally invariant derivative terms in a
Landau-Ginsburg theory such as $(\partial \phi /\partial
x)^4+(\partial \phi /\partial y)^4$.  All such terms are
irrelevant.  For a one-dimensional quantum fermion system the
dispersion relation is not usually exactly linear. However, the
additional terms in the Hamiltonian reflecting the non-linearity
involve at least two derivatives and hence are usually
irrelevant.  (Of course the situation changes if the linear term
vanishes.)  Once we have convinced ourselves that our critical
theory is scale invariant and Lorentz invariant it follows
immediately that it must be conformally invariant.

A quantum impurity problem cannot be invariant under the full
conformal group.  Indeed it is not even Lorentz invariant since a
special point is singled  out; or a special line, $x=0$ in the
space-time description.  The maximal symmetry that we could hope
for is the subgroup of the conformal group which leaves this line
fixed.  In the imaginary time formulation, introducing the
complex co-ordinate, $z=v\tau+ix$, the full conformal group is
all analytic transformations $z \to w(z)$.  The subgroup leaving
the line $x=0$, (the real axis) invariant is the set of
transformations for which $w(\tau ) \epsilon {\cal R}$.  Taylor
expanding the analytic function: \begin{equation} w(z) =
\sum_{n=1}^{\infty} a_nz^n\end{equation} we see that while, for a
general conformal transformation the $a_n$'s can be arbitrary
complex numbers, in the presence of a boundary they must all be
real.  Thus, the (infinite) number of symmetry generators is
reduced by a factor of 2.  This infinite set of symmetries
corresponds to time translations and rescalings of space and time
which may be performed independently at each space-time point.

We generally expect that a system which exhibits conformal
invariance far from a boundary will exhibit boundary conformal
invariance as the boundary is approached.  Such a system  could
be a  two-dimensional statistical system  at its bulk critical
temperature with a boundary. It could also be a one-dimensional
quantum system in a gapless phase with linear dispersion relation
in the presence of a quantum impurity.  Note that we have managed
to formulate the two-impurity Kondo model as such a system.  In
the left-right formalism the model is defined on the positive
$x$-axis with both impurities at $x=0$.  Of course, the universal
critical behavior only emerges in the scaling limit.  If we
calculate correlation functions we only expect universal behavior
when all points are well separated.  Furthermore all points must
be far from the boundary compared to microscopic scales.  Since
the ratios of distances from the boundary to distances between
the points remain as free parameters it is still possible to
observe critical behavior which is affected by the boundary.
Generally we can also make universal predictions when the points
are close to the boundary but far from each other.  However in
this case only exponents, not amplitudes are universal.  Again if
we assume invariance under time translations and global rescaling
of space or time the local invariance follows.

The specification of conformally invariant boundary conditions
can be conveniently formulated in terms of the finite-size
spectrum.  We consider a system defined on a cylinder of length
$l$ and circumference $\beta$ (ie. a quantum system at
temperature $T=1/\beta$) with some conformally invariant boundary
conditions $A$ and $B$ at the two ends.  Conformal invariance
implies that the spectrum can be specified in terms of the
conformal towers of the periodic left-moving system on an
interval of length $2l$.  The partition function can only be a
sum of partition functions for each conformal tower with a
multiplicity factor: \begin{equation} Z(l/\beta ) =
\sum_in^i_{AB} Z_i(l/\beta )\label{spec}\end{equation} Various
pairs of conformally invariant boundary conditions correspond to
various sets of integers $n^i_{AB}$.  It turns out that not all
possible choices of integers $n^i_{AB}$ correspond to a pair of
conformally invariant boundary conditions.  Cardy derived a set
of powerful consistency equations that these integers must
obey.\cite{Cardy}  In the particular case where the theory is a
product of several decoupled theories, as for all versions of the
Kondo problem, the set of conformal towers in Eq. (\ref{spec})
must be summed over all products of conformal towers from each
sector. For the two-impurity Kondo problem, there are four
sectors: the two isospins, total spin and Ising.  The number of
conformal towers in each sector is two for each isospin (labelled
by the isospin of the primary field, $i=0$ or $i=1/2$) three for
spin ($j=0,1/2,1$) and three for Ising (${\bf 1}$, $\sigma$ and
$\epsilon$).  Thus altogether there are 36 products of conformal
towers and 36 integers to specify.  For the free fermion spectrum
discussed in the previous section, six of these integers have the
value $1$ and the rest are $0$.  The $6$ products of conformal
towers that occur are given in Table I.

It turns out that a useful way of generating new conformally
invariant spectra from old ones is ``fusion''.\cite{Cardy,AL}
This corresponds to a particular mapping of each conformal tower
into a set of other ones.  The fusion rules are not obviously
related to boundary critical phenomena per se but come from the
bulk operator product expansion.  If we consider the OPE of two
bulk primary operators, ${\cal O}_i$ and ${\cal O}_j$ then it
will contain the complete set of primary fields ${\cal O}_k$ with
multiplicities $N^k_{ij}$.   The new spectrum is obtained by the
replacement: \begin{equation} n^i \to \sum_kN^i_{jk}n^k
\end{equation} for any fixed $j$. This procedure gives a new
solution of Cardy's consistency equations for any choice of the
conformal tower, $j$.  The fusion rule coefficients are a
property of each sector of the theory independently. They were
given for the Ising sector in Eq. (\ref{fusion}).

As a trivial example of a new conformally invariant spectrum
generated by fusion we consider fusion with the  $j=1$ field in
the spin sector, $\vec \phi$.  We see from the fusion rules that
this simply interchanges the $j=0$ and $j=1$ conformal towers.
The number of products of conformal towers remains six and is
given in Table II.  Notice that in this case the groundstate is
16-fold degenerate.  This spectrum corresponds to free fermions
with a $\pi /2$ phase shift.\cite{AL}  I.e. the Fermi level now
sits at one of the single-particle energy levels.  Since each of
the four species of Fermions (two channel times two spins) can
have this level filled or empty, this accounts for the 16-fold
degeneracy. In fact precisely this spectrum describes the Kondo
fixed point which occurs for large ferromagnetic inter-impurity
interaction.  We also note, that the same spectrum is obtained by
fusion with either the $\epsilon$ field in the Ising sector or
the $i_1=i_2=1/2$ field in the isospin sector.

The other possible fusion in the Ising sector is with the
$\sigma$ field. This gives the non-trivial critical point in the
two-impurity Kondo problem.  The resulting spectrum is given in
Table III.  In this case there are eight products of conformal
towers. Note that the same spectrum can be obtained by beginning
with the spectrum after $\vec \phi$-fusion, of Table II and then
performing $\sigma$-fusion. Thus the non-trivial fixed point is
symmetric with respect to the zero phase shift and $\pi /2$ phase
shift fixed points. We now wish to compare this in detail with
the NRG results.  This is done by multiplying together the
various conformal towers.  A general state will have energy
$E=(\pi v/l)(x+n)$. Here $x$ is the energy of the primary (or
highest weight) state and the integer $n$ is the descendent
level.  Both $x$ and $n$ are obtained by summing over the four
sectors.  The values of $x$ that occur for the eight products of
conformal towers are given in Table III.  We will measure
energies from that of the groundstate; thus $x-1/16$ occurs.  We
will content ourselves with working out the energies of all
states with $x+n-1/16 < 2$.  This implies $n=0$ or $1$ only.
Thus we only need know the first descendents in each conformal
tower.  Altogether there are $8$ different conformal towers in
the three inequivalent component theories ($k=1$ WZW, $k=2$ WZW
and Ising).  All first descendents of all $8$ conformal towers
are given in Table IV.  We now obtain the complete spectrum at
the non-trivial fixed point with $x+n-1/16 <2$ by taking either
the primary state in each sector or taking a first descendent in
one sector and primaries in the other three.  This gives the set
of states shown in Table V.  Recall that parity interchanges the
two isospins and also multiplies by $(-1)$ for the $\vec \phi$
conformal tower. The spectrum is symmetric between the two
isospins.  By taking symmetric or antisymmetric products of the
two isospins we obtain states of definite parity.

The general particle-hole symmetric Hamiltonian ((2.6) or (2.9),
with arbitrary coefficients) has three symmetries: total isospin
$\vec  I=\vec I_1+\vec I_2$, total spin, and (exchange) parity.
When the coupling constant $J_-=0$, a further symmetry develops,
and one can describe the eigenstates with an alternative set of
three quantum numbers: the two isospins, $i_1$ and $i_2$, and
total spin. This was actually the case in the NRG calculation to
which we compare the conformal field theory (CFT) spectrum.
Eigenstates within the same $i_1$, $i_2$ isospin-parity multiplets are
seen in Table V to be degenerate to better than one part in
$10\sp4$, an indication of the accuracy to which the symmetry
holds in the NRG calculations.\cite{foot1}

In Table V we express the lowest eigenstates via both sets of
quantum numbers. Columns one and two constitute a complete set:
the two isospins ($i_1$,$i_2$) and the total spin ($j$). To
facilitate comparison with the NRG, we have also resolved the
spectrum into states of definite $i$ (total isospin), $j$ (total
spin), and $P$, parity.  (Note that resolving states of definite
$i$ picks out states of definite symmetry with respect to
interchanging the two isospins and hence determines states of
definite parity.  In particular the $i=0$ part of an
$(i_1,i_2)=(1/2,1/2)$ multiplet
is antisymmetric and the $i=1$
part is symmetric.  An additional intrinsic odd parity must be
included for all states whose spin factor is in the $j=1$ conformal
tower.) Column five of Table V indicates $i$ and $P$ ($+\equiv$ even,
$-\equiv$ odd) in this
alternative formulation. Altogether we obtain $49$ multiplets
with definite $(i,j,P)$.  In the last column in Table V we give
the energies of the lowest  states of the same quantum numbers
obtained using the NRG.

Very briefly, numerical renormalization group
calculations\cite{NRG} iteratively diagonalize a Hamiltonian,
bringing in successive low-energy degrees of freedom until
asymptotically the ground state is reached. After each iteration,
the set of energy levels is ordered and truncated to keep only a
fixed number of the lowest-energy states. Truncation errors are
the main source of inaccuracy in the energy levels, and although
the truncation approximation is expected to be a good one for the
very lowest energies, the uncertainty (typically a few percent at
most) increases with increasing energy. For the results in column
six of Table V, approximately 1100 states were kept at each
iteration. The coupling constant $\nu J=0.18$ (c.f. Eqs. (2.6,
2.7)). For this value of $J$, the unstable fixed point is fully
realized by 14 iterations.

To determine the critical point, the ratio of RKKY coupling to
Kondo temperature was varied until the region of unchanging
energy levels was maximized. In practice, of course, numerically
one can never exactly sit at the critical point; since the
critical point occurs for a finite (probably irrational) value
of the coupling parameters, it is in practice only possible to
choose initial parameters such that the flows are asymptotically
close to the critical point.
And, in fact, no matter how carefully one adjusts the parameters,
at any iteration there are always some energy levels that are not
completely flat. Since for an inexact choice of parameters the critical
region does not extend all the way to $T=0$, one must choose a
temperature (iteration number) at which the approach to the critical
point is closest. These two facts, that one is never right at the
unstable point in parameter space, and that even with a very good
initial parameter choice, it is not clear which iteration to
choose to best represent the unstable fixed
point, thus introduce additional uncertainty in the numerical
energy values in Table V.

The agreement with NRG nonetheless seems very satisfactory. The
$49$ NRG states shown in Table V consist of numbers $1-47$,
$68-69$ in a consecutive numbering by increasing
energy.  The overall energy scale, set by $v$, is not determined
by the NRG so we have adjusted it to fit the CFT spectrum. In a
few cases the CFT predicts several multiplets with identical
quantum numbers and the same energy, so the matching with the NRG
levels involves some arbitrary choices. Otherwise, there are no
free parameters.  The agreement is to within about $2\%$ for the
first $17$ levels but gets progressively worse at higher
energies.  At the highest energies shown the disagreement is
about $5\%$ for most of the levels.\cite{foot1}
The two sets of multiplets
involving the ${1\over 2}'$ descendent in the spin sector (column
2 in table V) have unusually large disagreements with CFT
compared to other states at the same energy, $11\%$ and $19\%$
respectively.  It may be possible to understand this in terms of
the contribution of the leading irrelevant operator to finite-size
effects.\cite{Cardy2}  We regard Table V as rather convincing evidence
that we have identified the correct boundary CFT critical point for
this problem and that the NRG method is highly accurate.

It seems clear that the spectrum of Table V could not be obtained
from simply imposing linear boundary conditions, respecting
spin-rotation symmetry, on the free fermion fields, and filling up the
corresponding levels using Fermi statistics.  Note, for example, the
strange ratios of energy gaps: $3:4:7:8$ and  the peculiar 7-fold
degeneracy of the states with $x-1/16=1/2$.  Thus we refer to this as a
``non-Fermi liquid spectrum''.  Nonetheless, it is possible  to
describe this spectrum in terms of free Majorana Fermion field and their
spinor representations, using the alternative Ising $\times SO(7)_1$
bosonization scheme of Ref. (\onlinecite{AL2imp}). These fields are related
in a non-local way to the original conduction electron degrees of freedom. A
related physical picture   can also be found in Ref.'s
(\onlinecite{Sire,Gan}). We explain this hidden symmetry in  detail in Sec.
V.

\section{Operator Content} The operator content at the non-trivial
fixed point is determined by ``double fusion'' with $\sigma$.  This is
because we must effectively consider the spectrum on a strip with the
same boundary condition at each side.  The result in shown in Table
VI.  To understand the stability of the non-trivial fixed point we
must carefully consider all relevant and marginal boundary operators
($x\leq 1$) which are allowed by symmetry to appear in the Hamiltonian.
 (As usual in critical phenomena we expect that ``anything that can
happen will happen''.)  Apart from the identity operator, there are
seven other multiplets of relevant operators with $x=1/2$.  If we
assume that $\vec J$, $\vec I_1$ and $\vec I_2$ are all conserved then
only the seventh operator in Table VI, which we may write simply as
$\epsilon$, may appear in the Hamiltonian.  The $\tilde J_-$ coupling
constant of Eq. (\ref{Ham-bos}) breaks isospin down to the diagonal
subgroup.  This might seem to permit an element of the eighth operator
multiplet in Table VI: $(h_1^\dagger)^A(h_2)_{A}$.  Using Eq.
(\ref{daggerident}), this can be rewritten as
$(h_1)_A(h_2)_B\epsilon^{AB}$. Since we may regard the fields $h_1$
and $h_2$ as commuting, it follows that this operator is odd under
parity.  This conclusion also follows from the form of the Hamiltonian
in Eq. (\ref{Ham-bos}); since $\vec \phi$ is odd under parity, so must
be $(h_1^\dagger)^A(h_2)_{A}$.  Similarly the various marginal primary
operators in Table VI are not allowed by symmetry.  There are also
marginal descendent operators, the currents $\vec J$ and $\vec I_i$.
These are forbidden by spin and (diagonal) isospin symmetries.

Hence we reach the important conclusion that there is only one
relevant operator permitted, $\epsilon$.  It follows that the
non-trivial critical point can be reached, in general, by tuning one
parameter to its critical value.  We could vary any of the  coupling
constants in the problem to reach the critical point.  The physical
picture is clearest if we vary the RKKY coupling $K$, as explained in
the introduction.  We expect that varying $K$ away from its critical
value, $K_c$, drives the system to one of the stable fixed points with
either a $\pi /2$ phase shift in both channels or zero phase shift in
both channels.
 The coefficient of $\epsilon$ in the effective Hamiltoniam should be
proportional to $K-K_c$. In particular, it is important to note that
for the case of particle-hole symmetry, tuning such that $J_-=0$, that
is, setting even and odd-parity couplings equal, is not necessary for
obtaining the critical point as long as $K=K_c$.

There is a formal analogy between this renormalization group flow and
one that occurs in the classical Ising model.\cite{Cardy} Consider a
two dimensional Ising model at its critical temperature, defined on a
half-plane with free boundary conditions.  Then apply a weak magnetic
field at the boundary only, pointing up or down.  The free boundary
condition represents an unstable fixed point; applying the magnetic
field drives the system to the stable boundary condition with spins
pointing up or down at the boundary.
 The local magnetic field corresponds to the $\epsilon$ operator of
dimension $1/2$. The three different boundary conditions, spin up,
spin down and free are related to each other by fusion.  $\epsilon$
fusion takes spin up into spin down and vice versa; $\sigma$ fusion
takes spin up or down into free.

We may approach the critical point in various ways; one is by setting
$T=0$ and letting $K$ approach $K_c$; another is by setting $K=K_c$ and
letting $T$ go to zero.  If we tune $K$ to $K_c$, then the corrections to
scaling behaviour are governed by the leading irrelevant operator.  None of
the other primary operators in Table VI are allowed by symmetry.  The
leading irrelevant operator is the $x=3/2$ first descendent $ \epsilon '$.
It can be checked that the various other dimension $3/2$ operators are not
allowed by symmetry. In particular, $\vec J_{-1}\cdot \vec \phi$, the
leading irrelevant operator in the multi-channel, single-impurity Kondo
effect is not allowed by parity.

We may define the coefficient of the $ \epsilon '$ term in the
effective Hamiltonian as $1/\sqrt{T_K}$, where $T_K$ is the Kondo
temperature. Thus, at low temperatures and for $K$ close to $K_c$, we
must consider the following Hamiltonian: \begin{equation} H = H_{FP} +
\alpha {(K-K_c)\over \sqrt{T_K}}\epsilon + {1\over \sqrt{T_K}}\epsilon
', \end{equation} where $\alpha$ is a constant and $H_{FP}$ is the
Hamiltonian at the fixed point, with the non-trivial boundary
condition. We first consider the properties of $H_{FP}$ itself,
assuming that $K=K_c$ and ignoring the effects of the irrelevant
operator.  We then consider a non-zero $\epsilon '$ with $\epsilon$ still
zero, corresponding to approaching the critical point by setting $K=K_c$
and $T \to 0$.  Finally we consider $\epsilon$ and $\epsilon '$ both
non-zero corresponding to a general approach to the critical point.

A quantity which characterizes the critical point is the residual
impurity entropy.\cite{ALg}  This is defined by taking the length of
the system to $\infty$ first and then taking the temperature to zero.
In general the impurity entropy, i.e. the part which {\it does not}
scale with length, but rather goes to a constant as the length goes to
$\infty$, goes to a finite non-zero constant as $T\to 0$.  This
quantity is always the log of an integer for a Fermi liquid fixed
point, the integer simply being the degeneracy of the decoupled
impurity.  In the two-impurity Kondo problem, we expect zero
degeneracy at the stable fixed points since the impurity spins are
always in singlet states.  At a non-Fermi liquid fixed point, the
residual impurity entropy, $S(0)\equiv \ln g$ where $g$ need not be an
integer, in general. When the fixed point is obtained by fusion, a
general expression for $g$ has been derived in terms of the modular
S-matrix.  This gives: \begin{equation} g={S^{\bf I}_\sigma \over
S^{\bf I}_{\bf I}}=\sqrt{2}.\end{equation}  Note that $g$ {\it
decreases} under renormalization from the unstable critical point to
the stable ones, in accord with the ``$g$-theorem''.

Another striking signature of the non-trivial critical point is
provided by the zero-energy  one-particle S-matrix,
$S_{(1)}$.\cite{AL2ch}  This is the amplitude for a single electron to
scatter off the impurity and produce a single electron final state.
If the system exhibited Fermi liquid behavior then $S_{(1)}$ would
have unit amplitude  since all multi-particle final states would have
vanishing probability at zero energy.  Thus  $S_{(1)}$ would reduce to
a phase shift, $e^{2i\delta}$.  However, at a non-Fermi liquid fixed
point, the probability of inelastic scattering (one particle into
many) does not vanish at zero energy so  $S_{(1)}$ can have non-unit
amplitude without violating unitarity.  $1-|S_{(1)}|$ provides a
direct measurement of how ``non-Fermi liquid like'' the critical point
is.  It is a universal quantity characterizing the various fixed
points.  It is defined precisely by the asympotic behaviour of the
single electron Green's function.  In the absence of any impurity
couplings, this would have the non-interacting form: \begin{equation}
<\psi^{\alpha \dagger}_i(z)\psi _{\beta j}(z')>\to
{\delta^\alpha_\beta \delta_{ij}\over z-z'}\label{Gfree}\end{equation}
This dependence on spin indices will always be of the above form, due
to spin-rotation symmetry, so we will suppress it in what follows.
Ignoring the irrelevant, $\tilde J_-$ term in Eq. (\ref{H_K}), the
same is true of the dependence on the channel indices.  If $x$ and
$x'$ have the same sign, the form of Eq. (\ref{Gfree}) holds even in
the presence of impurity interactions.  This is a general feature of
conformally invariant boundary conditions.  On the other hand, when
the two fields at $x$ and $x'$ straddle the impurity, then different
asympotic behaviour can emerge, which depends on the universality
class of the boundary interactions.  The most general form consistent
with conformal invariance and spin and channel symmetries is simply:
\begin{equation} <\psi^{\alpha \dagger}_i(z)\psi _{\beta j}(z')>\to
{\delta^\alpha_\beta \delta_{ij}S_{(1)}\over
z-z'}\label{Gint}\end{equation} where $S_{(1)}$ is a universal
amplitude, which we can interpret as the S-matrix element for
one-particle into one-particle scattering.  In the non-interacting
case, $S_{(1)}=1$.  This should also be the case at the stable Fermi
liquid fixed point where the two impurities form a singlet.  On the
other hand, at the Fermi liquid fixed point where both impurities are
Kondo screened, $S_{(1)}=-1$, corresponding to a $\pi /2$ phase
shift.  As mentioned in Sec. I, $S_{(1)}$ must be real in all cases by
particle-hole symmetry.  $S_{(1)}$ is given by the matrix element of
the $\sigma$ operator (since this is the Ising factor of the fermion
operator, $\psi$) between the boundary state and the vacuum state.
Since the boundary state is obtained by fusion with the $\sigma$
primary field, this gives the result: \begin{equation} S_{(1)}=
{S^\sigma_\sigma S^{\bf I}_{\bf I}\over S^\sigma_{\bf I}S^{\bf
I}_\sigma}\end{equation} where $S^i_j$ is the modular S-matrix (which
has no obvious connections with the scattering matrix) of the Ising
model.  It turns out that $S^\sigma_\sigma =0$, while the other matrix
elements are non-zero.  Hence, we conclude that: \begin{equation}
S_{(1)}=0\end{equation} at the non-trivial fixed point.  This implies
that the scattering is entirely inelastic at zero energy.  Thus, is a
sense, this fixed point is as non-Fermi liquid like as possible.  The
fact that $S_{(1)}=0$ can be understood as a consequence of the
symmetry between the trivial fixed points with $S_{(1)}=1$ and
$S_{(1)}=-1$.  As mentioned in Sec. III, the non-trivial fixed point
can be reached by $\sigma$-fusion starting from {\it either} trivial
fixed point.  This implies that $S_{(1)}$ at the non-trivial fixed
point is proportional to its value at both trivial fixed points {\it
with the same constant of proportionality}.  Since $S_{(1)}$ has
opposite sign at the two trivial fixed points, it follows that it must
vanish at the non-trivial one.

If we relax the assumption that $\tilde J_-=0$, then the Green's
function need no longer be diagonal in channel indices. We may also
relax the assumption of particle-hole symmetry.  In general, at a
Fermi liquid fixed point we may write: \begin{equation} <\psi^{\alpha
\dagger}_i(z)\psi _{\beta j}(z')>\to {\delta^\alpha_\beta
S_{(1)ij}\over z-z'},\label{S1gen}\end{equation} where $ S_{(1)ij}$ is
a general unitary matrix.  It is the non-unitarity of $S_{(1)}$ which
signals non-Fermi liquid behaviour.

Ignoring all (relevant or irrelevant) perturbations about the critical
point, the impurity specific heat and the uniform impurity
susceptibility, which is the response of the system to a conserved
quantity, vanish. This occurs for the same reason as in the single
impurity case and we refer the reader to  Ref. (\onlinecite{AL}).  On
the other hand the staggered susceptibility does not vanish.  This is
defined in terms of the correlation function  of $\vec S_1-\vec S_2$.
This operator is odd under parity so we expect that it reduces to the
primary field $\vec \phi$ at the critical point.  This operator has
dimension $1/2$ giving a correlation function: \begin{equation} <[\vec
S_1-\vec S_2](\tau )\cdot [\vec S_1-\vec S_2](0)> \propto {1\over
|\tau |}\end{equation} This is the same as the impurity spin
correlation function in the overscreened two-channel single impurity
case mentioned above.  Fourier transforming at finite temperature and
analytically continuing to real frequency, we obtain [see third
reference in [\onlinecite{AL2ch}]] the staggered susceptibility.  The
real part behaves as: \begin{eqnarray} \hbox{Re}\chi_s(\omega ,T)
&\propto& {1\over T_K}\ln (T_K/T) \qquad (\omega <<T)\\ &\propto &
{1\over T_K}\ln (T_K/\omega ) \qquad (T<<\omega ).\end{eqnarray} The
imaginary part  behaves as: \begin{equation} \hbox{Im} \chi_s \propto
\tanh(\beta \omega /2). \end{equation}
 The electron spin operator with the same symmetry, $\psi_1^\dagger
\vec \sigma \psi_1-\psi_2^\dagger \vec \sigma \psi_2$ should also
couple to $\vec \phi$ and exhibit the same decay, close to the
boundary. The universal crossover function from bulk to boundary
behaviour could be calculated following the techniques of Ref.
(\onlinecite{AL2ch}).  Various other  operators exhibit the same
$|\tau |^{-1}$ decay due to coupling to  other $x=1/2$ boundary
operators in Table VI, $\epsilon$ and $(h_1)_A(h_2)_B$.
 As already remarked, the various interaction terms in the Hamiltonian
such as $\vec S_1\cdot \vec S_2$ couple to $\epsilon$ and so should
have this decay. Similarly, the spin-singlet, channel symmetric pair
operator, $(\psi_{\alpha 1}\psi_{\beta 2}+ \psi_{\alpha 2}\psi_{\beta
2})\epsilon^{\alpha \beta}$ should couple to $(h_1)_1(h_2)_1$.

Now we consider the effect of the leading irrelevant operator,
$(1/\sqrt{T_K})\epsilon '$, still assuming that $K=K_c$. We can
calculate physical properties by doing perturbation theory in
$1/\sqrt{T_K}$.  This is similar to the calculations for the
two-channel single impurity, $s=1/2$ Kondo effect, where the leading
irrelevant operator also has dimension $3/2$. However, in this case
the leading irrelevant operator is a Virasoro descendent, rather than
a primary.  Indeed, we may consider it to be the derivative of the
primary field, $\epsilon$: $\epsilon ' \approx d\epsilon /d\tau$.
Consequently the finite temperature two-point function has a different
form than in the two-channel case and we find that there is no
logarithmic singularity in the specific heat. I.e., \begin{equation}
c_{\hbox{imp}} \propto T/T_K.\end{equation} Note that the uniform
susceptibility [ie. the response to a field coupling to the  conserved
total spin of impurities plus conduction electrons] obtains no
contribution to second order in perturbation theory in the $\epsilon '$
term and therefore also has no logarithmic singularity.  This is again
unlike the two-channel case where another dimension 3/2 irrelevant
operator, $\vec J_{-1}\cdot \vec \phi$ is present in the effective
Hamiltonian.  It is absent in the two-impurity case by parity.  We
expect a finite impurity susceptibility as $T \to 0$, of $O(1/T_K)$,
coming from the dimension two irrelevant operator $\vec J^2$.

Let us now consider the behavior for $K$ slightly displaced from $K_c$.
 As mentioned, we expect the system to renormalize to one or the other
of the stable Fermi liquid fixed points.  At these fixed points the
impurity spins either form an inter-impurity singlet or else are
Kondo-screened.  There are various leading irrelevant operators of
dimension 2.  One such operator comes from the Ising sector; it is the
Ising energy-momentum tensor, $T_I$. When $K$ is very close to $K_c$
the system does not renormalize away from the unstable critical point
until a very low energy scale is reached of order $(K-K_c)^2/T_K$.
This plays the role of the lowest effective Kondo temperature at the
{\it stable} fixed point.  Thus we expect a term of the form
\begin{equation} \delta H = {T_K\over (K-K_c)^2}T_I
\label{T_I}\end{equation} in the effective Hamiltonian at the stable
fixed point.  The other leading irrelevant operators, such as $\vec
J^2$ in the spin sector, presumably have coefficients which remain
finite as $K \to K_c$.  We expect this because the delayed flow away
from the unstable critical point only occurs in the Ising sector.
The  uniform impurity susceptibility  goes to a constant as $T \to 0$
and can be determined by first order perturbation theory in the
leading irrelevant operators. It should remain finite as $K \to K_c$,
because it doesn't involve any Ising sector fields and is determined
by the coefficient of $\vec J^2$. On the other hand, the specific heat
does get a first order contribution from the irrelevant operator of
Eq. (\ref{T_I}).  Therefore, we expect a specific heat slope,
\begin{equation}\gamma \propto {T_K\over (K-K_c)^2}.\end{equation}  In
general, for finite $T$ and finite $K-K_c$, we expect the specific heat
to be given by some universal scaling function: \begin{equation}
c_{\hbox{imp}} = {T\over T_K}f\left({T\over T_K} , {K-K_c\over
T_K}\right),\end{equation} where the scaling function $f(x,y)$ has the
asymptotic behaviour: \begin{eqnarray} f(x,0) &\to & \hbox{constant}\
\ \  (x\to 0)\nonumber \\ f(0,y) &\to & {\hbox{constant}\over y^2}\ \
\  (y\to 0).\end{eqnarray} These scaling results are consistent with
those of the NRG, which, at $T=0$ find\cite{Jones} that $C_{imp}/T$
diverges as $1/(K-K_c)^2$, and  that the uniform susceptibility  is
non-singular as $K \to K_c$, in the presence of particle-hole symmetry.

We now wish to consider the effect of particle-hole symmetry breaking.
 The first type of P-H symmetry, defined in Sec. II corresponds to:
\begin{equation} \psi_{\alpha i} \to \epsilon_{\alpha \beta}\psi^{\beta
\dagger}_i\label{P-H}\end{equation}  In the bosonized representation, this
has the effect $(h_i)_1 \to (h_i)^{1 \dagger}$ or equivalently:
\begin{equation} (h_i)_A \to \epsilon_{AB}(h_i)_B  \end{equation} The
relevant particle-hole symmetry breaking interactions correspond to
potential scattering terms of the form: \begin{equation} \delta H =
V_e\psi^\dagger_e\psi_e+V_o\psi^\dagger_o\psi_o\label{pot-scatt}\end{equation}
(Here, the operators are evaluated at the origin in position space.)
In terms of the fields, $\psi_i$, this becomes: \begin{equation}
\delta H = {(V_e+V_o)\over
2}[\psi^\dagger_1\psi_1+\psi^\dagger_2\psi_2]+{(V_e-V_o)\over
2}[\psi^\dagger_1\psi_2+\psi^\dagger_2\psi_1]\end{equation} Upon
bosonizing, this gives: \begin{equation} \delta H \propto
(V_e+V_o)(I^z_1+I^z_2)+ \hbox{constant}\cdot
(V_e-V_o)\hbox{tr}(h_1)^\dagger \tau^3(h_2)\epsilon
\label{potscat}\end{equation} The first term produces the marginal
current operator.  In terms of an abelian total charge boson, $\phi$,
this is proportional to $\partial \phi /
 \partial x$.  We expect this to be exactly marginal and to produce a
line of fixed points with continuously varying phase shifts connected
to the non-trivial critical point.  This is similar to adding
potential scattering to the two-channel $s=1/2$ critical point.  We
expect the second operator to produce the relevant operator
$\hbox{tr}(h_1)^\dagger \tau^3(h_2)$ at the non-trivial critical
point.  There are various ways of seeing that this should occur.
 One is by observing that, after double fusion, $\epsilon$ can turn
into the identity operator.  Another is simply to observe that this
relevant operator transforms the same way as the second term in Eq.
(\ref{potscat}) under all the symmetries of the problem.  If $V_e=V_o$
and $\tilde J_-=0$, then the relevant operator is not allowed by
symmetry, since the original Hamiltonian has two independent
symmetries $I^3_1$ and $I^3_2$ whereas the relevant term only respects
the diagonal symmetry $I^3_1+I^3_2$.  If $\tilde J_-\neq 0$ then the
relevant term is allowed by symmetry even if $V_e=V_o$.  In this case
we expect it to appear.  Thus we conclude that potential scattering is
only relevant if $V_e\neq V_o$ or $\tilde J_-\neq 0$.
 We expect that the relevant potential scattering term will drive the
system to a Fermi liquid fixed point, corresponding to a unitary
$S_{(1)}$ of the general form of Eq. (\ref{S1gen}).  In the limit
where the relevant potential scattering term, $\hbox{tr}(h_1)^\dagger
\tau^3(h_2)$ has a very small coefficient, the flow to the stable,
Fermi liquid, fixed point occurs at a very low energy scale, so we
again expect that one of the leading irrelevant operators at the stable
fixed point will have a large coefficient of $O[T_K/(V_e-V_o)^2]$
(assuming that $\tilde J_-=0$).  This time, the large coupling
constant should occur in the isospin sector, rather than the Ising
sector, as above.  Thus the relevant potential scattering term should
produce a diverging specific heat coefficient $\propto
1/(V_e-V_o)^2$.  We do not expect uniform or staggered spin
susceptibilities to diverge as $V_e-V_0\to 0$ in this case, since the
spin operators do not involve the isospin fields.  However,
divergences should occur in charge and pair correlation functions.

Earlier NRG calculations\cite{baj4} are all in agreement with these
predictions of CFT regarding particle-hole symmetry breaking. When
$J_-=0$ and $V_e=V_o$, the two Fermi liquid fixed points and the
non-Fermi liquid critical non-trivial critical point are each extended
into a line a fixed points with continuously varying phase shift.
Potential scattering is thus irrelevant in this case. When $J_-\neq 0$
or $V_e \neq V_o$, all $T=0$ flows are to a single line of Fermi
liquid fixed points, connecting the antiferromagnetic RKKY singlet at
$\delta=0$ to the two-impurity Kondo point at $\delta=\pi/2$.
Potential scattering is relevant, and the non-trivial critical point
is washed out in this case.

Note that the first type of P-H symmetry is necessary to stabilize the
non-trivial fixed point.  The second type allows potential scattering with
$V_e=-V_o$.  This permits the relevant interaction discussed above [the
second term in Eq. (\ref{potscat})] so  the non-trivial
critical point should not occur in this case.

Finally, we show that an  anisotropy of the Kondo exchange
coupling constants $J_{\pm},J_m $ gives rise only
to  an irrelevant perturbation
of the non-Fermi-liquid fixed point. This shows that the
exchange-anisotropic version of the 2-impurity Kondo
Hamiltonian gives rise to the same  universal low temperature
physics as the isotropic one, described above.
More precisely, we consider an anisotropic Kondo
exchange coupling, such that the Hamiltonian
is only invariant under rotations about the z-axis
in $SU(2)$-spin space (so far, it  has been invariant under
rotations about {\it any } such rotation axis).
To show the irrelevance of the exchange anisotropy, we proceed
as in Ref.(\onlinecite{ALPC}). Since the exchange anisotropy breaks
the  total spin $SU(2)$ symmetry, some operators may occur
in the fixed point hamiltonian which were not allowed
in the isotropic case on symmetry grounds. We will now show
that the only additional such operators are irrelevant.
To see this we simply need to consult Table VI, of boundary
operators which exist at the non-Fermi-liquid fixed point:
Operators of spin $j=1/2$ are not allowed since
rotations about $2\pi $ about any axis in $SU(2)$-spin
space leaves the anisotropic hamiltonian invariant,
but multiplies these operators by $(-1)$. Thus we are
left with  the spin $j=1$ operators in Table VI:
Amongst those, the only
  relevant or marginal new operators  (scaling
dimension $x\leq 1$)
that could  occur in the anisotropic case
are (i): $ \phi^z $, with quantum numbers $(i_1,i_2,j,Is) =$
$(0,0,1,{\bf 1})$, (ii): $\phi^z \epsilon$, with quantum
numbers $(0,0,1,\epsilon)$, as well as  (iii) $ (h^{\dagger}_1)^A
 (h_2)_A \phi^z $ with quantum numbers
$(1/2,1/2,1,{\bf 1}) $  and (iv): the spin-descendant $J^z $, with
quantum numbers $(0,0,1',{\bf 1})$. However, operators (i) and
 (ii)  are odd under parity, and are thus not not allowed
to occur since the microscopic (exchange anisotropic)
 Hamiltonian is parity invariant. Operator (iii)
changes sign under rotations by $\pi $ about the $x-$
or $y-$ axis in spin space whereas the microscopic
Hamiltonian does not. Hence (iii) cannot occur.
Operator (iv) is
 odd under time
reversal [this can be seen e.g. from Eq.(\ref{Ham-bos})]
and  thus cannot occur either ( the latter
does, in any case, not renormalize, and could not destroy the
non-Fermi-liquid fixed point for that reason).
Hence we conclude that no relevant operators can occur
at the fixed point even  in the presence of exchange anisotropy.

\section{Hidden $SO(7)$ Symmetry}
An interesting feature of the spectrum at the non-trivial fixed
point, discussed in Sec. III, is various apparently ``accidental''
degeneracies.  For instance the second and third conformal tower in
Table III have the same value of $x-1/16=3/8$ although they have
different values of $(i_1,i_2,j)$. This suggests the presence of
some type of  symmetry  which is higher than the spin and
isospin symmetries discussed so far. A related observation was made in
Ref. (\onlinecite{baj3}): { there,  an additional conserved quantity
was noticed
at the non-trivial fixed point, namely, $\int dx\left(\vec J_e-\vec
J_o\right)$ where, \begin{equation} \vec J_e-\vec J_o\equiv
\psi^\dagger_e{\vec \sigma \over 2}\psi_e-\psi^\dagger_o{\vec
\sigma\over 2}\psi_o= \psi^\dagger_1{\vec \sigma\over
2}\psi_2+\psi^\dagger_2{\vec \sigma \over
2}\psi_1.\label{Je-Jo}\end{equation} In fact, a higher symmetry
{ has been found in
Ref. (\onlinecite{AL2imp})
 and corresponds
to the group $SO(7)$ which contains $SU(2)^{i_1}\times
SU(2)^{i_2}\times SU(2)^j$ as a subgroup.
[ This is in fact a conformal embedding, $SO(7)_1 \to$
$SU(2)_1^{i_1}\times SU(2)_1^{i_2}\times SU(2)_2^j$.]}  In this
section we spell out this symmetry in more detail.

We emphasize at the outset that this symmetry is rather different
and more mysterious than the isospin and spin symmetries discussed
so far in that it does not become manifest for any choice of the
Kondo couplings in Eq. (\ref{H_K}), except the trivial case where
all couplings are zero.
In this case the full symmetry is actually
$SO(8)$.  This follows because we have $4$  species
of complex or Dirac
fermions, equivalent to $8$ Hermitean or Majorana fermions.  This
$SO(8)$ symmetry is also present at the Kondo screened fixed point
{ (where the RKKY interaction is strongly ferromagnetic)
since the phase shift boundary condition does not break the symmetry
between the $8$ Majorana fermions.  Remarkably, an $SO(7)$ subgroup
of this $SO(8)$ also remains at the non-trivial fixed point.  The
key to understanding this symmetry is the realization that the
non-trivial critical point is related to the free fermion one by
fusion in the Ising sector.  Thus what we must do is to represent
the free fermions as a product of an Ising model and another sector
of maximal possible symmetry.  This symmetry will be preserved
under fusion in the Ising sector. The maximally symmetric
representation corresponds to a conformal embedding or alternative
non-abelian bosonization scheme where the $8$ free fermions are
represented by an $SO(7)_1$ Kac-Moody conformal field theory
together with  an Ising conformal field theory.
  This preserves the value of the
conformal anomaly or Virasoro central charge parameter c,
 since the KM algebra $SO(7)_1$ has $c=7/2$ and the Ising model,
$c=1/2$:
 \begin{equation}c= 4=  7/2 +
1/2.\end{equation} There are actually two inequivalent ways of
constructing a conformal embedding depending on the $SO(7)$
representations into which the $8$ Majorana fermions, transforming
under the vector representation of $SO(8)$, decompose.  The trivial
embedding corresponds to:
 \begin{eqnarray}
  SO(8)_1 &  \to &  SO(7)_1 \times Is \nonumber \\
 8  & \to &  (7,{\bf 1}) + (1,\epsilon).
 \end{eqnarray}  [Here and in the following we label $SO(n)$
representations by their dimensions and conformal towers by the
representation of their highest weight state.]
This simply takes advantage of
the fact that the Ising model is itself equivalent to a free Majorana
fermion.  We may define the   $SO(7)$ transformation to
rotate $7$ of the fermions and leave the $8^{\hbox{th}}$ invariant.
However, this is {\it not} the appropriate embedding; ie. this Ising
model does not appear in the representation of the fermions in the same
way as the one which we discussed above.  Rather the correct
representation corresponds to an alternative embedding in which:
 \begin{eqnarray}
  SO(8)_1 &  \to &  SO(7)_1 \times Is \nonumber \\
 8  & \to &  (8,\sigma).
 \end{eqnarray}
  Here, the $8$ on the right-hand-side of this equation is
the $8$-dimensional spinor representation of $SO(7)$.   [ See
any book on Lie groups, e.g. Ref.(\onlinecite{georgi}),
regarding the properties of the orthogonal groups.]
 In general, the  scaling   dimension, $x$,
of a  Kac-Moody  primary field transforming  under
a representation  $\rho$ of a
group is given by\cite{WZW}:
 \begin{equation}x_{\rho}={C_{\rho} \over C_A+k},
\end{equation}
 where $C_{\rho}$ is the quadratic Casimir invariant
for the representation $\rho$, $C_A$ is the quadratic Casimir for the
adjoint representation and $k$ is the Kac-Moody central
  extension (level).  For
the spinor (8), vector (7) and adjoint (21) representations of
$SO(7)$, the Casimirs are:
\begin{eqnarray}
C_7 &=& 3\nonumber \\
C_8 &=& 21/8\nonumber \\
C_{21}&=& 5.\end{eqnarray}
Taking $k=1$, we obtain   for the scaling dimensions
\begin{eqnarray}
x_7&=&1/2\nonumber \\
x_8&=& 7/16.\end{eqnarray}
We see that a consistent conformal embedding represents the $8$
Majorana fermions, $\chi_A$, of  $SO(8)_1$,
 in terms of the $SO(7)_1$ spinor field,
$S_A$, ($A=1,2,3,...8.$) and the Ising order parameter field,
$\sigma$: \begin{equation}
\chi_A \propto S_A\cdot \sigma .\end{equation}
Since $x_\sigma = 1/16$, the dimension adds up correctly to $1/2$
for a free fermion.  Comparing to our previous bosonization formula of
Eq. (\ref{Isingbos}) we note that the Ising factor, $\sigma$, is
common to both formulae and that the $SO(7)$ spinor field has replaced
the product of $SU(2)$ spin and isospin fields.
The free fermion spectrum of Table I can be
rewritten in terms of $SO(7)_1\times \hbox{Ising}$ conformal towers.
 We may  now use the  conformal embedding
of $SO(7)_1$  conformal towers [labeled by dimensions] into
 $SU^{i_1}(2)_{1}\times SU^{i_2}(2)_{1}$
$ \times SU^{j}(2)_2$
  conformal towers [labeled by angular momenta]:
\begin{eqnarray}
7 &\to& (1/2,1/2;0) + (0,0;1)\nonumber \\
8 &\to& (1/2,0;1/2) + (0,1/2;1/2),\end{eqnarray}
 (Note that both KM algebras have the same conformal anomaly parameter:
 $ c = 7/2
= 1 + 1 + 3/2$.)
 From the equation above and TABLE I,  we find
for the combinations of
conformal towers of
the free fermion spectrum: $(SO(7)_1,\hbox{Ising})=
(1,{\bf 1}),
(8,\sigma ), (7,\epsilon )$.

It is now immediately evident that the $SO(7)$ symmetry will be
preserved  in  any new spectrum obtained from this one by fusion {\it
with an Ising field only}.  In particular, the
combinations of conformal towers  at the
non-trivial fixed point are given in this notation by:
 $(SO(7)_1,\hbox{Ising})=
(1,\sigma ),
(8, {\bf 1} ), (8, \epsilon ), (7,\sigma )$.  Note that the two-fold
degenerate sets of representations in Table III at $x-1/16=3/8,
1/2$ and $7/8$ collapse to  single representations of $SO(7)$ in all
three cases.  The conformal tower at $x-1/16=1$ in Table III
actually corresponds to a descendent of $(1,\sigma )$ from the
$SO(7)$ point of view.  We may also simplify Table V by constructing
the spectrum up to $x-1/16 <2$ in this new basis.
The first
descendents of the $SO(7)$ conformal towers are given by:
\begin{eqnarray}
SO(7)_1  {\rm -primary} & \to &
SO(7)_1 \ {\rm -descendant} \nonumber \\
1 &\to& 21'\nonumber \\
7 &\to& 7', 35'\nonumber \\
8 &\to& 8', 48'.\end{eqnarray}
We find that all states in Table V can be grouped into $SO(7)$
representations and this explains all ``accidental'' degeneracies.
(Of course other degeneracies persist which arise from the fact that
the spacing of levels of all conformal towers in any conformal field
theory is $\delta x=1$.)

We remark that there is  an isomorphism\cite{fuchs} of the
fusion rules of the
$SO(7)_1$ and the Ising conformal field theories,
corresponding to the following mapping of
$SO(7)_1$ and Ising conformal towers:
 \begin{eqnarray}
SO(7)_1 &\leftrightarrow& {\rm Ising}  \nonumber \\
1 &\leftrightarrow& {\bf 1} \nonumber \\
8 &\leftrightarrow& \sigma \nonumber \\
7 &\leftrightarrow& \epsilon  \end{eqnarray}
 Note that since   the free fermion
spectrum  corresponds to a sum of products of the three pairs of
identified conformal towers, namely $ (SO(7)_1, Is)=$
$ (1,{\bf 1})$, $(8, \sigma )$, $(7,
\epsilon )$, we can go from the free fermion to
non-trivial fixed point by fusion either with $\sigma$ in the Ising
sector or, equivalently, with 8 in the $SO(7)_1$ sector.

The decomposition of the adjoint representation
of $SO(7)$, under  which the $SO(7)_1$ current [a KM descendant
of the $SO(7)_1$ identity operator] transforms, gives information
about the additional conserved quantities:
\begin{eqnarray}
SO(7)_1 & \to & SU^{i_1}(2)_1\times SU^{i_2}(2)_1\times
SU^{j}(2)_2
\nonumber \\
 21' &  \to &
(1',0;0)+(0,1';0)+(0,0;1')+(1/2,1/2;1).
\end{eqnarray}
   At  the zero coupling, infinite coupling and   non-trivial fixed
points, there  are 21 (Hermitean) conserved quantities with these quantum
numbers.  The first 9 clearly correspond to the currents, $\vec I_1$, $\vec
I_2$
and $\vec J$.  The remaining 12 operators are uniquely identified by
their quantum numbers as $\psi^\dagger_1\vec \sigma \psi_2$,
$\psi_1\sigma^2\vec \sigma \psi_2$ together with the hermitean conjugates of
these operators.  We see that {\it one}  linear combination of these
corresponds
to the extra conserved quantity identified previously in the NRG work, Eq.
(\ref{Je-Jo}).

The 28 fermion bilinears that can be formed from 8 Majorana fermions,
 which transform under the adjoint representation of $SO(8)$
and represent the $SO(8)_1$ KM current operator, decompose
into the 21 and 7 dimensional representation of $SO(7)$.
Thus under  the conformal embedding $SO(8)_1 \to
SO(7)_1 \times Is$, 21  of the
 fermion bilinears become
 the $SO(7)$ current
and the remaining 7 become the product of the
vector (7)  and $\epsilon$  primaries.
  It can be seen that both
types of bilinears appear in the Kondo interactions.  As stated
above, these interactions do not manifestly preserve the $SO(7)$
symmmetry for any non-zero value of the couplings; rather this
symmetry is ``dynamically restored'' at the non-trivial critical
point.

  This  (non-abelian)  bosonization scheme of the 4 species of
Dirac Fermions in terms of an $SO(7)_1$ Wess-Zumino-Witten
theory and an Ising model
 [first pointed out in Ref. (\onlinecite{AL2imp}) ],
 appears to be closely related to  a
physical picture for
the 2-impurity Kondo problem  proposed more
recently by Sire et al.\cite{Sire} and Gan.\cite{Gan}
 In these papers,
 the fermions are first transformed by ordinary abelian
bosonization.  Certain linear combinations of bosons are defined and
then it is observed that  a subset of the boundary
operators can be re-expressed in terms
of some different fermions, a transformation sometimes referred to
as ``refermionization''.
  For a special, anisotropic, value of the
Kondo couplings, up to irrelevant operators, it is found that only 1
of the new Majorana fermion couples to the impurity spins, the other 7
completely decoupling.
 This theory  has an obvious $SO(7)$ symmetry.
This procedure apparently gives  an explicit realization of the
$SO(7)$ bosonization scheme, with the added bonus that the $SO(7)$
symmetry at the non-trivial critical point becomes manifest.

\section{Discrepancy with Other Calculations}  Finite temperature Monte
Carlo (MC) calculations by Fye and Hirsch.\cite{Fye} have so far discovered
no evidence for the non-trivial critical point discussed here. It was
suggested that the non-trivial critical point may be an artifact of the
energy-independent coupling constant approximation used in the NRG work.
Furthermore, NRG calculations by Sakai et al.\cite{Sakai} on the related
2-impurity Anderson model, while reproducing the results of Ref.
(\onlinecite{Jones}) in the energy (and parity) independent coupling
constant case, did not find such a critical point using energy-dependent
coupling constants or a ``parity-splitting'' term.  In this section we
discuss the apparent discrepancy between these calculations and the ones
presented in the present paper.

Our most important point  was derived in detail in Sec. II
and Sec. V; it is the existence of two different types of P-H symmetry,
only one of which leads to a non-trivial fixed point. As explained in Sec.
II, when the first type of  particle-hole symmetry is maintained, some sort
of phase transition as a function of inter-impurity coupling, $K$, is
inevitable.\cite{Millis}  The reason is that for large antiferromagnetic
$K$ there is no phase shift and for large ferromagnetic $K$ the phase shift
is $\pi /2$ in both channels.  Furthermore, because of this type of
particle-hole symmetry, the phase shifts, if they are well-defined, can
only be $0$ or $\pi /2$.  Both of these phases are absolutely stable (no
relevant or marginal operators). Thus they must be separated by (at least)
one point which is not in either phase.  In principle, this could either be
a first order transition point or a critical point. This argument seems to us
very general and convincing. It should not, for instance, depend on ignoring
energy-dependence of the coupling constants, discussed in Sec. II. It only
relies on the rather well-established local Fermi liquid picture of the
single-impurity Kondo effect developed by Nozi\`eres and others.  The
statement that the phase shift is $\pi /2$ for some range of parameters has a
precise mathematical meaning.  It is a statement about the asymptotic
behaviour of the electron Green's function, discussed in Sec. V.  If the
Green's function did not have this behaviour for a range of ferromagnetic $K$
this would be rather shocking and would presumably indicate a breakdown of
the standard theory of the single-impurity Kondo effect.  Likewise for a
range of antiferromagnetic $K$.  We think that measuring the quantity,
$S_{(1)}$, governing the asymptotic behaviour of the electron Green's
function, defined in the previous section, would be the easiest way to verify
the existence of the non-trivial critical point.  It ought to be possible to
show that $S_{(1)}=1$ for sufficiently large antiferromagetic inter-impurity
coupling and $S_{(1)}=-1$ for sufficiently large ferromagnetic inter-impurity
coupling.  A numerical study of how and where it passes between
$1$ and $-1$ would give information about the non-trivial
critical point.  We emphasize that $S_{(1)}\neq \pm 1$
corresponds to non-Fermi liquid behaviour.  At $T=0$ some sort of
phase transition must exist between these two Fermi liquid
cases.  An analysis of scaling with temperature will be required
to study this transition.

Of course, this argument says nothing about the nature of the
phase transition separating the two stable phases.  It could, for
example, be first order, in which case there would not be a
critical point.  To determine the nature of the phase transition
requires other methods.  The NRG and CFT methods provide
complementary approaches, which taken together, suggest rather
clearly that the two phases are separated by the non-trivial
critical point discussed in this paper.  The CFT approach allows a systematic
classification of possible critical points, given the assumption of
conformal invariance at the critical point and the various symmetries of
the problem.  The NRG results allow a determination of which of the possible
critical points actually occurs.  Having identified a candidate
critical point, and shown that it does occur in a particular
microscopic formulation of the model (corresponding to the NRG
with energy-independent coupling constants), we can then
determine whether it will be stable under changes in the
microscopic Hamiltonian using the RG.  The complete set of
relevant and marginal operators at the proposed critical point are
those in Table VI, together with the marginal current operators,
$\vec J$, $\vec I_i$.  If the microscopic Hamiltonian has
the first type of particle-hole symmetry (together with parity and at least a
$U(1)$ subgroup of the spin-rotation symmetry) then all but one of these
operators are forbidden to appear in the effective Hamiltonian of Eq.
(\ref{H_K}).  The presence of the first type of particle-hole symmetry
implies\cite{Jones,baj3,ALPC} the diagonal isospin symmetry, $\vec I$.  This
diagonal isospin symmetry together with spin symmetry [or at least a $U(1)$
subgroup of it] forbids all these operators except $\epsilon$ and  an
element of the eighth operator in Table VI, which is forbidden by
parity. (See the beginning of Sec. V.)  Thus we reach the crucial
conclusion that the non-trivial critical point can be reached by
adjusting only one parameter, the coupling constant corresponding
to  $\epsilon$ in the effective Hamiltonian, assuming the first
type of particle-hole symmetry.  By varying the inter-impurity coupling, $K$,
or any other parameter, so as to pass from the inter-impurity singlet phase to
the Kondo-screened phase, we can make this coupling constant pass
through zero.  Thus this is a conventional critical point; it is
reached by varying only one parameter, assuming  the
first type of particle-hole symmetry.

This analysis indicates that allowing energy-dependent coupling
constants should not make any difference, provided that the first type of
particle-hole symmetry is maintained.  No lowering of symmetry
ensues from allowing energy-dependent coupling constants, as long as the
first type of P-H symmetry is maintained, so no additional operators are
permitted in the effective Hamiltonian describing the critical point.  Of
course, the actual location of the critical point, as a function of the
various Kondo and RKKY couplings can change.

The NRG work on the Anderson model\cite{Sakai} considers several
different Hamiltonians.  The model with energy-independent
coupling constants, reduces, in the Kondo limit of the Anderson
model, to the same model studied previously\cite{Jones}; in this
case apparently the same non-trivial critical point is observed.
The other models considered, with energy-dependent coupling
constants or a ``parity-splitting'' term {\it do not have the first type of
particle-hole symmetry}. Thus, by the discussion, in Sec. II, we do not
expect them to exhibit the non-trivial critical point, and indeed they do
not.  Some of these models do have the second type of particle-hole
symmetry.  This symmetry is not enough
preserve the critical point as we discussed in Sec. II, V.  However, it
should be possible to study models with energy-dependent coupling constants
which do not break the first type of particle-hole symmetry, and which
therefore should exhibit the non-trivial fixed point.   This corresponds to
choosing couplings  $N_{e,o}(E)$ [corresponding essentially to $W_{0,1}(E)$
in the notation of Ref. (\onlinecite{Sakai})] which are non-trivial even
functions of energy, $E$, for both parities, [$p=0,1$].  (The other
parameters in the Hamiltonian must also be chosen to preserve particle-hole
symmetry.)  Further calculations with such models would be valuable in
settling this controversy.

Most of the QMC work of Ref. (\onlinecite{Fye}) was done on the
one-dimensional tight-binding model at half-filling with
$\delta$-function Kondo interactions, of the form:
\begin{equation}J[\psi^\dagger (R/2) \vec \sigma \psi (R/2)\cdot
\vec S_1+\psi^\dagger (-R/2) \vec \sigma \psi (-R/2)\cdot
\vec S_2],\label{HRS}\end{equation} where $R=na$, $a$ being the lattice
spacing. [In the case $n$ odd, the origin is midway between 2 sites.]  The
various cases, $n=1,2,4,8$ were studied.  We see from Eq. (\ref{alpha})
that the first type of P-H symmetry occurs for $n$ even and the second for
$n$ odd.  A systematic search for the non-trivial critical point was only
made in the case $n=1$, where it is not expected to occur!  The reason for
this was connected with the fact that these authors did not put in an
inter-impurity coupling by hand.  They expect (based on weak-coupling
perturbation theory) that the RKKY coupling will be ferromagnetic for $n$
even and antiferromagnetic for $n$ odd.  The NRG work of Ref.
(\onlinecite{Jones}) suggests that the non-trivial critical point should
occur for antiferromagnetic inter-impurity coupling, of $O(T_K)$. From
another point of view, it is quite easy to see that these models are
generally in the Kondo-screening phase at strong coupling, so if a
transition is to be encountered they had better be in the inter-impurity
singlet phase at weak coupling.  A necessary condition for this is
presumably an antiferromagnetic RKKY coupling.

There is another potential problem with the even $n$ models.  In this
case $N_e(0)=0$, [see Eqs. (\ref{eodef}), (\ref{Neodef}]  so the even
channel  has only irrelevant Kondo couplings.  If we assume the even channel
decouples, we obtain a single-channel Kondo effect with a spin-1 impurity
for ferromagnetic inter-impurity coupling.  This leads to a Fermi liquid
fixed point, with an underscreened $s=1/2$ effective impurity.  However, we
expect that higher orders of perturbation theory will generate a non-zero
Kondo couplings to the even channel at $E=0$.  Such a coupling to the
leftover $s=1/2$ impurity would be relevant, provided that it is
antiferromagnetic.  However, if it is ferromagnetic, the underscreened
Fermi liquid fixed point would be stable and the symmetry argument given
above for the existence of the non-trivial critical point may fail.

Thus it seems quite likely that
the non-trivial critical point will not occur for any value of $n$ and
$J$ in this model. However, it should be possible to see the critical point
by generalizing the model somewhat, while maintaining the desired
symmetry. It may be sufficient to take an even $n$ model (say
$n=2$) and add a direct inter-impurity coupling, $K$.  To avoid the
potential difficulty mentioned in the previous paragraph, it may also be
necessary to modify the Kondo couplings somewhat while preserving the
first type of P-H symmetry. It can be seen that this
symmetry occurs at half-filling provided that there is site-parity, ie.
reflection symmetry about a site as occurs for $n$ even.  The second type
occurs if there is link-parity, ie. reflection about a link.  One
example would be to make the replacement in Eq. (\ref{HRS}):
\begin{eqnarray} \psi(R/2) &\to &  v_0\psi (0)+iv_1\psi (a)\nonumber
\\\psi(-R/2) &\to &v_0\psi (0)+iv_1\psi (-a).\end{eqnarray}   This
gives, in wave-vector space:
\begin{equation} v(k) = v_0+iv_1e^{iak}.\end{equation}
 The first
type of P-H symmetry holds provided that the $v_i$'s are real.
For $v_0$ and $v_1$ both non-zero, $N_{e}(0)$ and $N_{o}(E)$ are both
non-zero. However, we find that the RKKY coupling is always ferromagnetic
for all the models of this type that we have considered with the first type
of P-H symmetry.  Thus it may be necessary to add a direct  inter-impurity
coupling which is varied to pass between the two stable phases.

It may still be difficult to see the non-trivial critical point by studying
the staggered susceptibility since it only shows a logarithmic divergence.
It may be necessary to choose couplings very close to the critical value
and very low temperatures to see the anomalous behavior.  As discussed
above, the single-particle Green's function would likely exhibit a much
clearer signal of the critical point.

In conclusion, we expect that further numerical work, along the lines of
Refs. (\onlinecite{Sakai}) and (\onlinecite{Fye}) should be able to see the
critical point provided that:
\newline 1) the first type of particle-hole symmetry is maintained,
\newline 2) a parameter is varied to pass from the inter-impurity singlet to
Kondo screened phase and
\newline 3) sufficiently low temperatures and finely tuned parameters are
obtained.
\newline The last condition should be much easier to achieve if the
single-particle Green's function is measured instead of the staggered
susceptibility.

\centerline{\bf ACKNOWLEDGEMENTS}

We thank J. Gan, A. Millis and C. Varma for very helpful discussions. The
research of I.A. is supported in part by NSERC of Canada; A.W.W.L. is a
Fellow of the A.P. Sloan Foundation.

\begin{table} \caption{Conformal towers occurring for free
fermions with the boundary conditions of Eq.
(\protect{\ref{freebc}}).} \label{tab:freect}
\begin{tabular}{lllll} $i_1$ & $i_2$ & $j$ & Ising & $x$ \\
\tableline 0 &   0 &     0 & {\bf 1}& 0\\ 1/2 & 0 & 1/2 & $\sigma$ &
1/2 \\ 0 & 1/2 & 1/2 & $\sigma$ & 1/2\\ 0 & 0 & 1 & $\epsilon$
&1\\ 1/2 & 1/2 & 1 & {\bf 1} & 1\\ 1/2 & 1/2 & 0 & $\epsilon$ &
1\\ \end{tabular} \end{table}

\begin{table} \caption{Conformal towers occurring for free
fermions with a $\pi /2$ phase shift, corresponding to $\vec
\phi$ [or $\epsilon$ or $(1/2,1/2)$] fusion.}
\label{tab:ph_sh_ct} \begin{tabular}{lllll} $i_1$ & $i_2$ & $j$ &
Ising & $x$ \\ \tableline 0 & 0 & 1 & {\bf 1} &1/2\\ 1/2 & 0 &
1/2 & $\sigma$ & 1/2 \\ 0 & 1/2 & 1/2 & $\sigma$ & 1/2\\ 0 &     0
&        0 & $\epsilon$ & 1/2\\ 1/2 & 1/2 & 0 & {\bf 1} & 1/2\\ 1/2 &
1/2 & 1 & $\epsilon$ & 3/2\\ \end{tabular} \end{table}

\begin{table} \caption{Conformal towers occurring after $\sigma$
fusion.} \label{tab:sigma_ct} \begin{tabular}{lllll} $i_1$ &
$i_2$ & $j$ & Ising & $x-{1\over 16}$ \\ \tableline 0 &  0 &     0 &
$\sigma$ & 0\\ 1/2 & 0 & 1/2 & {\bf 1} & 3/8 \\ 0 & 1/2 & 1/2 &
{\bf 1} & 3/8\\ 0 & 0 & 1 & $\sigma$ &1/2\\ 1/2 & 1/2 & 0 &
$\sigma$ & 1/2\\ 1/2 & 0 & 1/2 & $\epsilon$ & 7/8 \\ 0 & 1/2 &
1/2 & $\epsilon$ & 7/8 \\ 1/2 & 1/2 & 1 & $\sigma$ & 1\\
\end{tabular} \end{table}

\begin{table} \caption{All first descendents of all conformal
towers.  The subscript in the first row labels the Kac-Moody
level, $k$. $SU(2)$ descendents are labelled by their spin.  All
descendents will be marked $'$.} \label{tab:descendents}
\begin{tabular}{llllllll}
 $(0)_1$ & $\left({1\over 2}\right)_1$ & $(0)_2$ & $\left({1\over
2}\right)_2$ & $(1)_2$ & {\bf 1} & $\sigma$ & $\epsilon$ \\ \\
\tableline \\ $1'$ & ${1\over 2}'$ & $1'$ & ${1\over
2}'$,${3\over 2}'$ & $0'$, $1'$ & -- & $\sigma '$ & $\epsilon
'$\\ \end{tabular} \end{table}

\begin{table} \caption{Spectrum at the non-trivial fixed point.}
\label{tab:spec} \begin{tabular}{lllllr} $(i_1,i_2)$  & $j$ &
Ising & $x-{1\over 16}$ & $i^P$& $E_{\hbox{NRG}}$\\ \tableline
(0,0)  &         0 & $\sigma$ & 0 & $0^+$ & .00000 \\ \tableline
$({1\over 2},0)$, $(0,{1\over 2})$  & ${1\over 2}$ & {\bf 1} &
${3\over 8}$ & ${1\over 2}^+$ & .37761 \\ &&&&${1\over 2}^-$&
.37764\\ \tableline (0,0)&1&$\sigma$&${1\over 2}$&$0^-$&.50454 \\
\tableline $({1\over 2},{1\over 2})$&0&$\sigma$&${1\over
2}$&$0^-$&.50724 \\ &&&&$1^+$&.50726 \\ \tableline $({1\over
2},0)$, $(0,{1\over 2})$&${1\over 2}$&$\epsilon$&${7\over
8}$&${1\over 2}^-$&.88696\\ &&&& ${1\over 2}^+$&.88702\\
\tableline (0,0)&$1'$&$\sigma$&1&$0^+$&.99952 \\ \tableline
$({1\over 2},{1\over 2})$&1&$\sigma$&1&$0^+$&1.00296\\
&&&&$1^-$&1.00298\\ \tableline
$(1',0)$,$(0,1')$&0&$\sigma$&1&$1^-$&1.00642\\ &&&&$1^+$&1.00644\\
\tableline (0,0)&0&$\sigma '$&1&$0^+$&1.01078\\ \tableline
$({1\over 2},0)$, $(0,{1\over 2})$&${3\over 2}'$&{\bf 1}&$1{3\over
8}$&${1\over 2}^+$&1.39056\\ &&&&${1\over 2}^-$&1.39066 \\
\tableline
 $({1\over 2}',0)$,$(0,{1\over 2}')$&${1\over 2}$&{\bf
1}&$1{3\over 8}$&${1\over 2}^+$&1.39321\\ &&&&${1\over
2}^-$&1.39329\\ \tableline $(1',{1\over 2})$,$({1\over
2},1')$&${1\over 2}$&{\bf 1}&$1{3\over 8}$&${3\over
2}^-$&1.39699\\ &&&&${1\over 2}^+$&1.39700\\ &&&&${1\over
2}^-$&1.39709\\ &&&&${3\over 2}^+$&1.39716\\ \tableline $({1\over
2},0)$,$(0,{1\over 2})$&${1\over 2}'$&{\bf 1}&$1{3\over
8}$&${1\over 2}^+$&1.52680\\ &&&&${1\over 2}^-$&1.52687\\
\tableline (0,0)&$1'$&$\sigma$&$1{1\over 2}$&$0^-$&1.50437\\
\tableline $({1\over 2}',{1\over 2})$,$({1\over 2},{1\over
2}')$&0&$\sigma$&$1{1\over 2}$&$0^-$&1.51090\\ &&&&$1^+$&1.51094\\
&&&&$1^-$&1.57554\\ &&&&$0^+$&1.57556\\ \tableline
(0,0)&$0'$&$\sigma$&$1{1\over 2}$&$0^-$&1.56562\\ \tableline
$({1\over 2},{1\over 2})$&$1'$&$\sigma$&$1{1\over
2}$&$0^-$&1.56892\\ &&&&$1^+$&1.56895\\ \tableline
$(1',0)$,$(0,1')$&1&$\sigma$&$1{1\over 2}$&$1^+$&1.57224\\
&&&&$1^-$&1.57226\\ \tableline (0,0)&1&$\sigma '$&$1{1\over
2}$&$0^-$&1.61452\\ \tableline $({1\over 2},{1\over
2})$&0&$\sigma '$&$1{1\over 2}$&$0^-$&1.62305\\
&&&&$1^+$&1.62312\\ \tableline $({1\over 2},0)$,$(0,{1\over
2})$&${1\over 2}$&$\epsilon '$&$1{7\over 8}$&${1\over
2}^-$&1.92017\\ &&&&${1\over 2}^+$&1.92029\\ \tableline
$(0,{1\over 2})$,$({1\over 2},0)$&${3\over
2}'$&$\epsilon$&$1{7\over 8}$&${1\over 2}^-$&1.97512\\
&&&&${1\over 2}^+$&1.97514\\ \tableline $({1\over
2}',0)$,$(0,{1\over 2}')$&${1\over 2}$&$\epsilon$&$1{7\over
8}$&${1\over 2}^-$&1.97817\\ &&&&${1\over 2}^+$&1.97824\\
\tableline $({1\over 2},1')$,$(1',{1\over 2})$&${1\over
2}$&$\epsilon$&$1{7\over 8}$&${1\over 2}^-$&1.98224\\
&&&&${3\over 2}^+$&1.98225\\ &&&&${1\over 2}^+$&1.98230\\
&&&&${3\over 2}^-$&1.98236\\ \tableline $({1\over
2},0)$,$(0,{1\over 2})$&${1\over 2}'$&$\epsilon $&$1{7\over
8}$&${1\over 2}^-$&2.23938\\ &&&&${1\over 2}^+$&2.23947\\
 \end{tabular} \end{table}

\begin{table} \caption{Operator content at the non-trivial
critical point.} \label{tab:opcnt} \begin{tabular}{lllll} $i_1$ &
$i_2$ & $j$ & Ising & $x$ \\ \tableline 0 &      0 &     0 & {\bf 1}& 0\\
1/2 & 0 & 1/2 & $\sigma$ & 1/2 \\ 1/2 & 0 & 1/2 & $\sigma$ & 1/2
\\ 0 & 1/2 & 1/2 & $\sigma$ & 1/2\\ 0 & 1/2 & 1/2 & $\sigma$ &
1/2\\ 0&0&1&{\bf 1}&1/2\\ 0&0&0&$\epsilon$&1/2\\ 1/2&1/2&0&{\bf
1}&1/2\\ 0 & 0 & 1 & $\epsilon$ &1\\ 1/2 & 1/2 & 1 & {\bf 1} & 1\\
1/2 & 1/2 & 0 & $\epsilon$ & 1\\ 1/2&1/2&1&$\epsilon$&3/2\\
\end{tabular} \end{table}


\begin{references}
\bibitem {foot1} Of the states listed in Table V, the
(1/2',1/2)(1/2,1/2') grouping is unique in that it is the only one
in the table whose degeneracy is not a consequence of the explicit
two-isospin plus total-spin symmetries. Instead, it relies on more
of the ``hidden" symmetries of the CFT. (More such groupings will
occur at higher energies.) Thus this particular degeneracy will not
be exact for a finite system, but should only set in asymptotically
as the system size increases.
\end{references}
 \end{document}